\documentstyle[10pt,emulateapj]{article}
 
\input{epsf}

\journalid{}{}
\articleid{}{}
\begin{document}
\def\lax    {\ifmmode{_<\atop^{\sim}}\else{${_<\atop^{\sim}}$}\fi}
\def\gax    {\ifmmode{_>\atop^{\sim}}\else{${_>\atop^{\sim}}$}\fi}
\def\gtorder{\mathrel{\raise.3ex\hbox{$>$}\mkern-14mu
             \lower0.6ex\hbox{$\sim$}}}
\def\ltorder{\mathrel{\raise.3ex\hbox{$<$}\mkern-14mu
             \lower0.6ex\hbox{$\sim$}}}
 
\long\def\***#1{{\sc #1}}
 
\title{XMM-Newton observations of the M31 northern disk: Properties of selected X-ray sources and diffuse emission.}

\author{Sergey Trudolyubov\altaffilmark{1,2,3}, Oleg Kotov\altaffilmark{2,4}, 
William Priedhorsky\altaffilmark{2}, France Cordova\altaffilmark{1}, and 
Keith Mason\altaffilmark{5}}

\altaffiltext{1}{University of California, Riverside, CA 92507}

\altaffiltext{2}{Los Alamos National Laboratory, Los Alamos, NM 87545}

\altaffiltext{3}{Space Research Institute, Russian Academy of Sciences, 
Profsoyuznaya 84/32, Moscow, 117810 Russia}

\altaffiltext{4}{Harvard-Smithsonian Center for Astrophysics, Cambridge, MA}

\altaffiltext{5}{Mullard Space Science Laboratory, University College 
London, UK}

\begin{abstract}
We present the results of XMM-Newton survey of the northern part of the disk of M31. 
The results of a spectral and timing analysis of the thirty seven brightest sources 
are presented. Combining the results of X-ray analysis with available data at 
other wavelengths, we were able to classify $\sim 50 \%$, or 19 out of 37 sources.

Two sources in our sample were previously unknown: the hard X-ray source 
XMMU J004415.8+413057 and a transient supersoft source XMMU J004414.1+412206. We 
report the discovery of possible X-ray pulsations from the source 
XMMU J004415.8+413057 with a period of 197 s. The spectral and timing properties of 
XMMU J004415.8+413057 make it first accreting X-ray pulsar candidate detected in 
M31.

We found six X-ray sources to be coincident with M31 globular cluster (GC) candidates. 
The spectral properties of GC sources were found to be similar to that of the bright 
Galactic low mass X-ray binaries located in the bulge and globular clusters. The 
comparison of the X-ray properties of GC sources with optical properties of their host 
globular clusters has shown that the brightest sources with luminosity, $L_{\rm X}$ 
above $\sim 10^{38}$ ergs s$^{-1}$ tend to reside in more metal poor clusters. 

Three X-ray sources were identified with SNR based on their X-ray spectra and positional 
coincidence with SNR candidates from optical and radio surveys. We found five bright X-ray 
sources to coincide with Galactic foreground stars. Two of them demonstrate a remarkable 
variability of X-ray flux on a time scale of individual {\em XMM-Newton} observations. 
Two X-ray sources coincide with radio sources and are probably distant radio galaxies/AGN. 

The properties of the remaining 18 bright X-ray sources detected in our survey are 
consistent with AGN in the background of M31 and X-ray binaries belonging to M31. 
Many of them show high values of low-energy absorption, which combined with extreme 
faintness of their possible optical counterparts makes them an obscured AGN candidates. 

We report on the first unambiguous detection of the soft unresolved X-ray emission 
from the disk of M31. The unresolved emission follows the pattern of the spiral arms 
and can be traced up to distance of $\sim 0.5\arcdeg$ ($\sim 7$ kpc at 760 kpc) from 
the center of the galaxy. The spectrum of the unresolved emission shows dominant soft 
thermal component which can be fitted with a $\sim 0.3$ keV optically thin thermal 
plasma emission models. We suggest that significant part of this diffuse soft X-ray 
component may represent hot diffuse gas in the spiral arms of M31 and emission from 
normal stars in the disk of M31. 
\\
\\
{\em Subject headings:} galaxies: individual (M31) --- X-rays: galaxies --- X-rays: stars
\end{abstract}

\section{INTRODUCTION}
The Andromeda Galaxy (M31), the closest spiral galaxy to our own (760 kpc; van der Bergh 2000), 
is a unique object for the study of X-ray astronomy. M31 is in many respects similar to the 
Milky Way and even called its ``twin sister''. M31 hosts hundreds of X-ray sources, which 
are observed at a nearly uniform distance, and due to the favorable orientation of the M31, 
they are less obscured by interstellar gas and dust than those in the Galaxy. M31 was observed 
extensively with {\em Einstein}, {\em ROSAT}, {\em Chandra} and {\em XMM-Newton} missions, 
detected hundreds of X-ray sources, identified with different types of X-ray emitting objects 
(\cite{TF91}; \cite{Primini93}; Supper et al. 1997, 2001; \cite{Kong02}; \cite{Shirey01}). 
Although the central part of M31 was subject to deep X-ray surveys, the outer disk of the 
galaxy was covered by observations with relatively low sensitivity (\cite{Williams03}).  

In this paper we present the results from the deep {\em XMM-Newton} observations of the northern 
parts of the disk of M31. We concentrate on the spectral and timing analysis of a selected sample 
of 37 bright X-ray sources. We also report on the first unambiguous detection of the diffuse 
X-ray emission associated with disk of M31. 

\section{OBSERVATIONS AND DATA ANALYSIS}
\begin{figure}
\hbox{
\begin{minipage}{9.0cm}
\epsfxsize=9.0cm
\epsffile{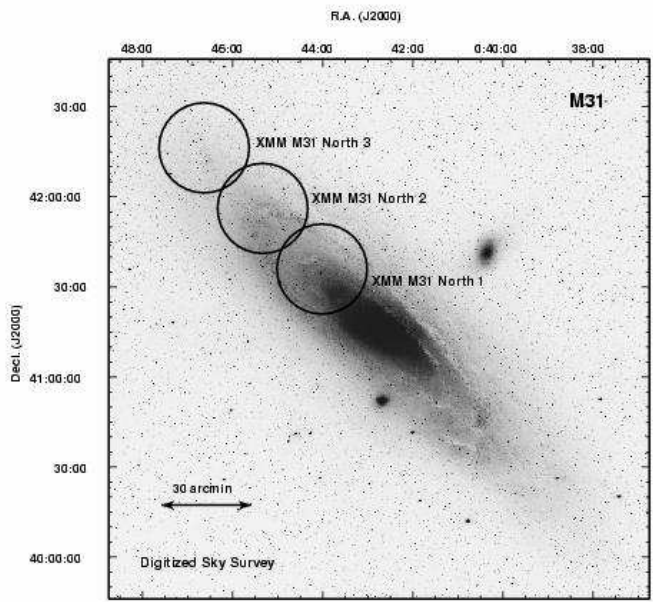}
\end{minipage}
\begin{minipage}{9.0cm}
\epsfxsize=9.0cm
\epsffile{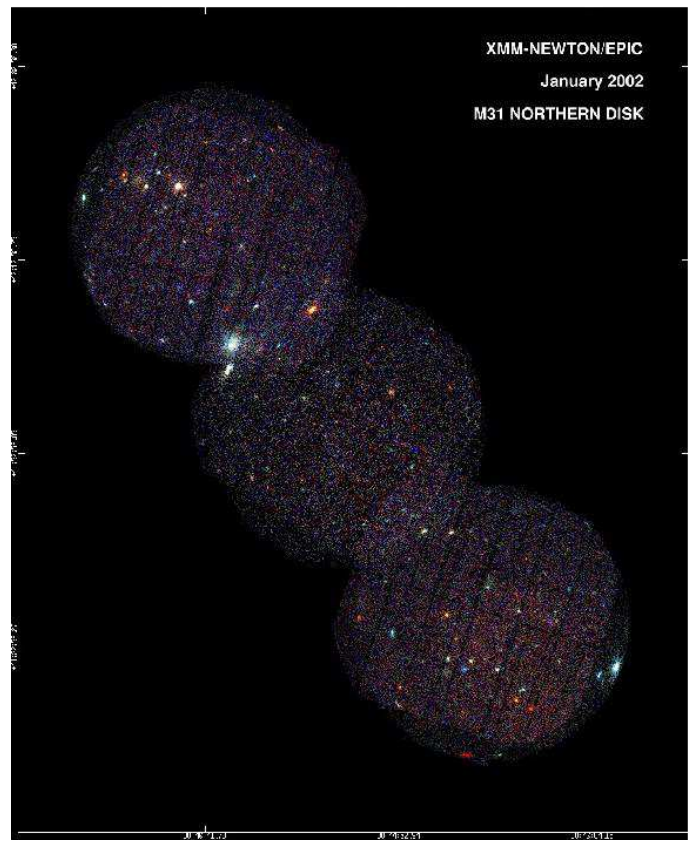}
\end{minipage}
}
\caption{\small {\em Left:} Optical image of M31 from Digitized Sky Survey with {\em XMM-Newton} 
FOV shown as circles for each of the M31 North 1, North 2 and North 3 fields from Table 
\ref{obslog}. {\em Right:} A detailed three-color combined {\em XMM-Newton}/EPIC X-ray 
image of the northern disk of M31. The red, green and blue intensities correspond to 
logarithmically scaled counts in the $0.3 - 1.0$, $1.0 - 2.0$ and $2.0 - 7.0$ keV energy 
bands. The image was constructed with $2\arcsec$ pixels.\label{image_optical_xray}}  
\end{figure}

Three regions of the northern part of galactic disk of M31 were observed with {\em XMM-Newton} 
on January 5 (Obs. $\#1$)(\cite{xmm_circ02}), January 26 (Obs. $\#2$), and June 29, 2002 (Obs. 
$\#3$) as a part of the Guaranteed Time Program (PI: K. O. Mason) (Table \ref{obslog}, Figure 
\ref{image_optical_xray}). The 2002, January 6 {\em XMM-Newton} observation of the central part 
of M31 (\cite{dipper02}) (Obs. $\#4$) provides additional coverage of the inner disk. In the 
following analysis we use the data from three European Photon Imaging Camera (EPIC) instruments: 
two EPIC MOS detectors (\cite{Turner01}) and the EPIC-pn detector (\cite{Strueder01}). In the first 
three observations EPIC instruments were operated in the {\em full window} mode 
($30\arcmin$ FOV) with the {\em medium} optical blocking filter. In the Obs. $\#4$ a combination 
of the {\em full window} mode and {\em thin} optical blocking filter was used.
 
We reduced EPIC data with the {\em XMM-Newton} Science Analysis System (SAS v 5.3)
\footnote{See http://xmm.vilspa.esa.es/user}. We performed standard screening of the 
EPIC data to exclude time intervals with high background levels. 

Images in celestial coordinates with a pixel size of 2$\arcsec$ have been accumulated in the 
$0.3 - 1.0$, $1.0 - 2.0$, $2.0 - 7.0$ and $0.3 - 7.0$ keV energy bands for the EPIC-MOS1, MOS2 and 
pn detectors. We produced a three-color image of the northern disk regions of M31 combining the 
images in the $0.3 - 1.0$ (soft), $1.0 - 2.0$ (medium) and $2.0 - 7.0$ keV (hard) energy bands and 
using red, green and blue color scales to represent the X-ray intensities in these bands 
(Fig. \ref{image_optical_xray}). The spectral energy distribution of the source defines its color 
in this image. The sources with soft X-ray spectra (i.e. supersoft sources, thermal supernova 
remnants and Galactic foreground stars) appear in red, while sources with hard spectra are blue. 

X-ray sources were detected with program based on a wavelet decomposition algorithm, set at a 
$4\sigma$ threshold. For our current analysis, we expect error in the source position determination 
to be dominated by residual systematic error of the order $2 - 5 \arcsec$. We corrected the count 
rates of the sources for the vignetting of the XMM telescope, based on the Current Calibration Files 
provided with SAS. 

We studied the spectral and timing properties of all X-ray sources in our sample. Each source in our 
sample provided between 300 and 18000 counts in EPIC detectors allowing high quality spectroscopic and 
timing analysis.

To generate lightcurves and spectra of X-ray sources, we used elliptical extraction 
regions with semi-axes size of $\sim 20 - 80 \arcsec$ (depending on the distance of the 
source from the telescope axis) and subtracted as background the spectrum of adjacent 
source-free regions with subsequent normalization by a ratio of the detector areas. We 
used data in the $0.3 - 10$ keV energy band because of the uncertainties in the calibration 
of the EPIC instruments outside this range. All fluxes and luminosities derived from spectral 
analysis apply to this band, unless specified otherwise. For the sources with soft (Galactic 
foreground stars, SNR candidates) and supersoft X-ray spectra we considered only the $0.3 - 3.0$ 
and $0.2 - 1.0$ keV spectral ranges, since their flux was negligible above 3.0 and 1.0 keV 
respectively. We used spectral response files generated by XMM SAS tasks. Spectra were grouped 
to contain a minimum of 20 counts per spectral bin in order to allow $\chi^{2}$ statistics and 
fit to analytic models using the XSPEC v.11\footnote{http://heasarc.gsfc.nasa.gov/docs/xanadu/xspec/index.html} 
fitting package (\cite{arnaud96}). EPIC-pn, MOS1 and MOS2 data were fitted simultaneously, but 
with normalizations varying independently.

Fourier power density spectra (PDS) were produced using the lightcurves in the 
$0.3 - 7.0$ keV energy band. Then we performed folding analysis in the vicinity of the 
frequency peaks identified from PDS. We used standard XANADU/XRONOS v.5
\footnote{http://heasarc.gsfc.nasa.gov/docs/xanadu/xronos/xronos.html} tasks to perform 
analysis of the timing properties of bright X-ray sources.

In the following analysis we assume a source distance of 760 kpc (van den Bergh 2000).

\section{SELECTED X-RAY SOURCES: IDENTIFICATION AND CLASSIFICATION}
\begin{figure}
\hbox{
\begin{minipage}{9.0cm}
\epsfxsize=9.0cm
\epsffile{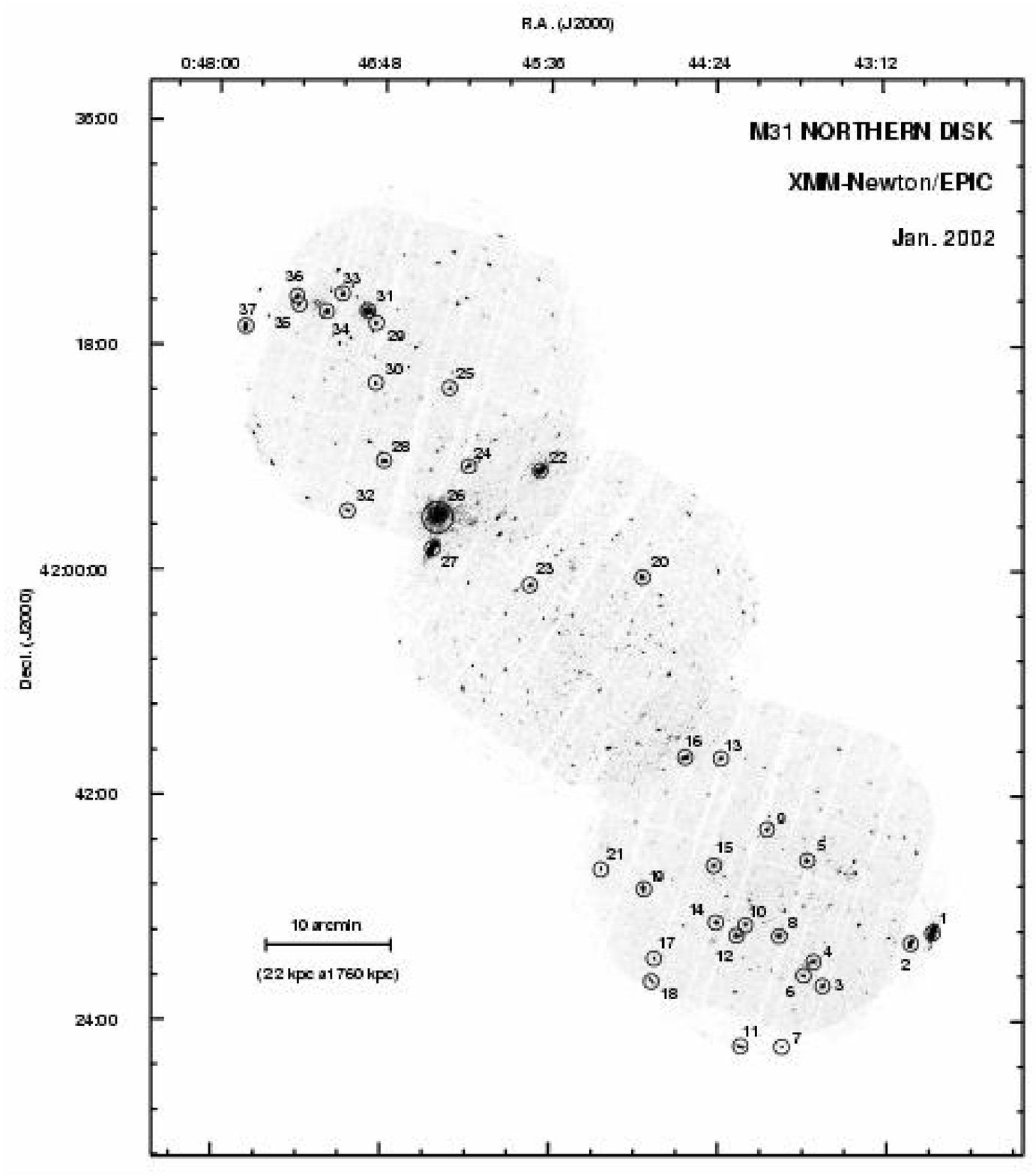}
\end{minipage}
\begin{minipage}{9.0cm}
\epsfxsize=9.0cm
\epsffile{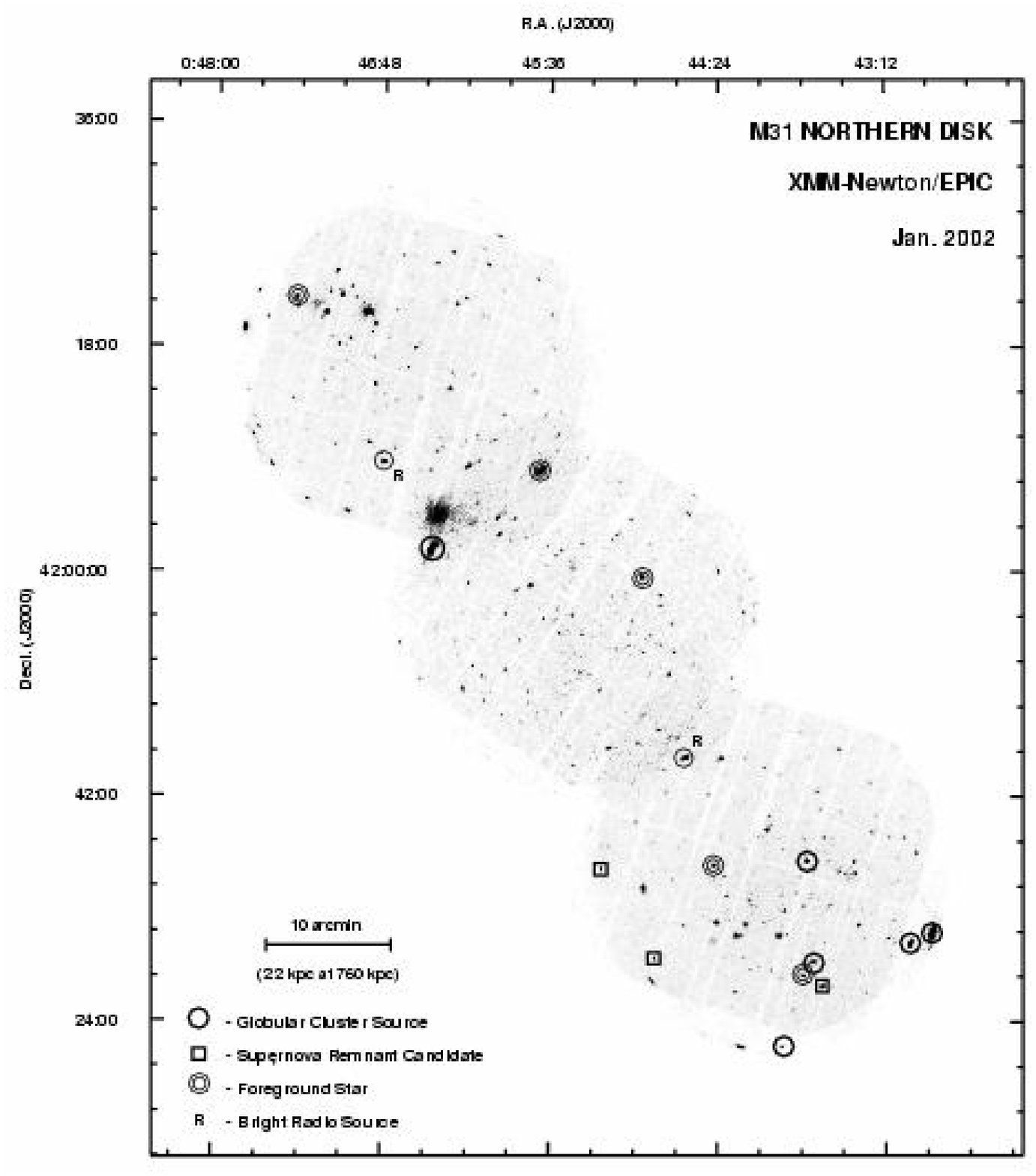}
\end{minipage}
}
\caption{\small {\em Left:} Combined {\em XMM-Newton}/EPIC image of the M31 North1, North2 and 
North3 disk fields (Table \ref{obslog}). Positions of the selected bright X-ray sources 
detected with XMM-Newton/EPIC cameras are marked with small circles. Source labels 
correspond to the numbering in Table \ref{source_ID}. {\em Right:} Combined 
{\em XMM-Newton}/EPIC image of the M31 North1, North2 and North3 disk fields with 
optical/radio identifications of the selected bright XMM sources from Table 
\ref{source_ID}. \label{image_epic_all}}
\end{figure}

Using the procedure described above, we detected about 300 X-ray point sources in the 
three {\em XMM} M31 northern disk fields. The complete X-ray source catalog will be 
presented in a follow-up paper. Here we study a sample of 37 sources selected on the 
basis of their brightness; each individual source was required to have more than 300 counts 
in the EPIC. The information on the positions and identifications of the selected X-ray sources 
is shown in Table \ref{source_ID}. The X-ray images of the northern disk regions of M31 with 
source positions marked are shown in Fig. \ref{image_epic_all}.   

We searched for optical, infrared and radio counterparts to the bright XMM sources in 
the northern disk of M31 using the existing catalogs and images from the CTIO/KPNO Local 
Group Survey (LGS) (\cite{Massey01}) and the Second Generation Digitized Sky Survey. We 
varied the search radius based on both the accuracy of the catalogs and localization 
errors of {\em XMM} sources (upper limit of $5\arcsec$). We used the following catalogs 
and corresponding search radii:

\noindent i) {\em X-ray sources:} the {\em ROSAT}/PSPC catalog of sources 
in the field of M31 (Supper et al. 1997, 2001) (SHP97,SHP01) -- search 
radius specified by position accuracy for each individual source. All but 
five bright {\em XMM} sources were detected in the {\em ROSAT}/PSPC survey 
of M31 (Table \ref{source_ID}). 

\noindent ii) {\em Globular cluster candidates:} the Bologna catalog 
(\cite{Bo87}), the catalog by Magnier (1994), and the HST globular cluster 
candidate catalog (\cite{Barmby01}) -- with search radius of $5 \arcsec$. 

\noindent iii) {\em Supernova remnant candidates:} the lists by Braun \& 
Walterbos (1993) and Magnier et al. (1995) -- $10\arcsec$ search radius.

\noindent iv) {\em Stellar objects (Galactic foreground stars/background AGN):} 
the catalogs of stellar objects by Magnier et al. (1992) and Haiman et al. 
(1994) and SIMBAD -- $3\arcsec$ search radius.

\noindent v) {\em Radio sources:} VLA All-sky Survey Catalog
\footnote{http://www.cv.nrao.edu/nvss/} (\cite{nvss})-- $5\arcsec$ search radius.

Inspection of optical images to search for uncatalogued bright star-like objects 
revealed no additional foreground star candidates. X-ray source $\# 33$, however, is very likely 
a background galaxy based on the shape of its counterparts in the LGS and DSS images.

\section{BRIGHT X-RAY SOURCES DISCOVERED WITH XMM-NEWTON}
\subsection{X-ray pulsar candidate XMMU J004415.8+413057}
\begin{figure}
\epsfxsize=18cm
\epsffile{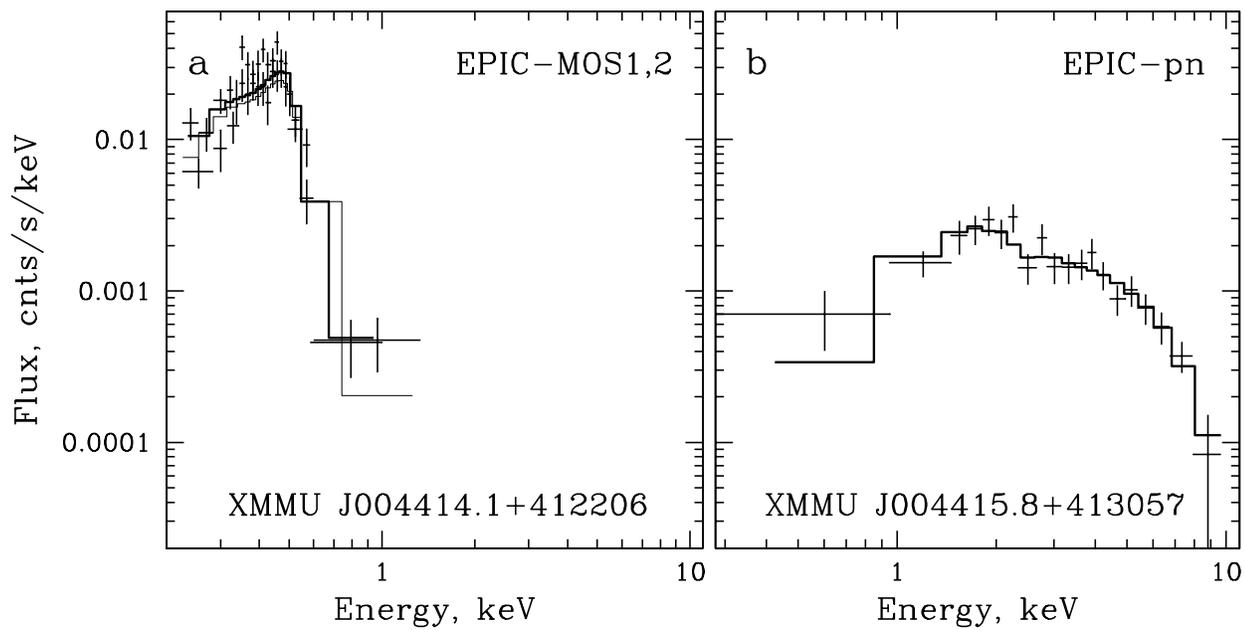}
\caption{\small Count spectra of bright transient sources. EPIC data, $0.3 - 10$ keV energy 
range. {\em (a)} Transient supersoft source XMMU J004414.1+412206, EPIC-MOS1 and MOS2 
data, fit by the absorbed black body radiation model. The fits to EPIC-MOS1 and MOS2 
data are shown with {\em thick} and {\em thin} histograms. {\em (b)} Hard X-ray transient 
source XMMU J004415.8+413057, EPIC-pn data, fit by the absorbed simple power law model. 
\label{spec_TR_fig}}
\end{figure}

The X-ray source XMMU J004415.8+413057 was discovered with {\em XMM-Newton} in the 2002 January 5 
observation (\cite{xmm_circ02}). As the source falls into a CCD gap in both EPIC-MOS detectors, we 
present the analysis of the source properties based on the EPIC-pn data alone. The spectrum of XMMU 
J004415.8+413057 is remarkably hard and heavily absorbed at low energies (Fig. \ref{spec_TR_fig}). 
The analytic approximation of the spectrum with an absorbed simple power law model gives a photon 
index of 1.4 and requires an absorbing column of $1.3 \times 10^{22}$ cm$^{-2}$ (Table 
\ref{spec_par_TR}; Fig. \ref{spec_TR_fig}). The corresponding absorbed $0.3 - 10$ keV luminosity of 
the source was $\sim 9 \times 10^{36}$ ergs s$^{-1}$. 

The X-ray source XMMU J004415.8+413057 shows a significant variability on a time scales of 
several years. It was not detected in the previous observations with {\em Einstein} and {\em ROSAT}. 
Using the data of archival 1996, July 7 deep {\em ROSAT}/HRC observation, and assuming the same 
spectral shape as measured with {\em XMM}, we estimate a $2 \sigma$ upper limit to the source 
luminosity of $\sim 2 \times 10^{36}$ ergs s$^{-1}$. This implies that XMMU J004415.8+413057 was 
at least a factor of 4 fainter during {\em ROSAT} observation in July 1996.

\begin{figure}
\epsfxsize=18cm
\epsffile{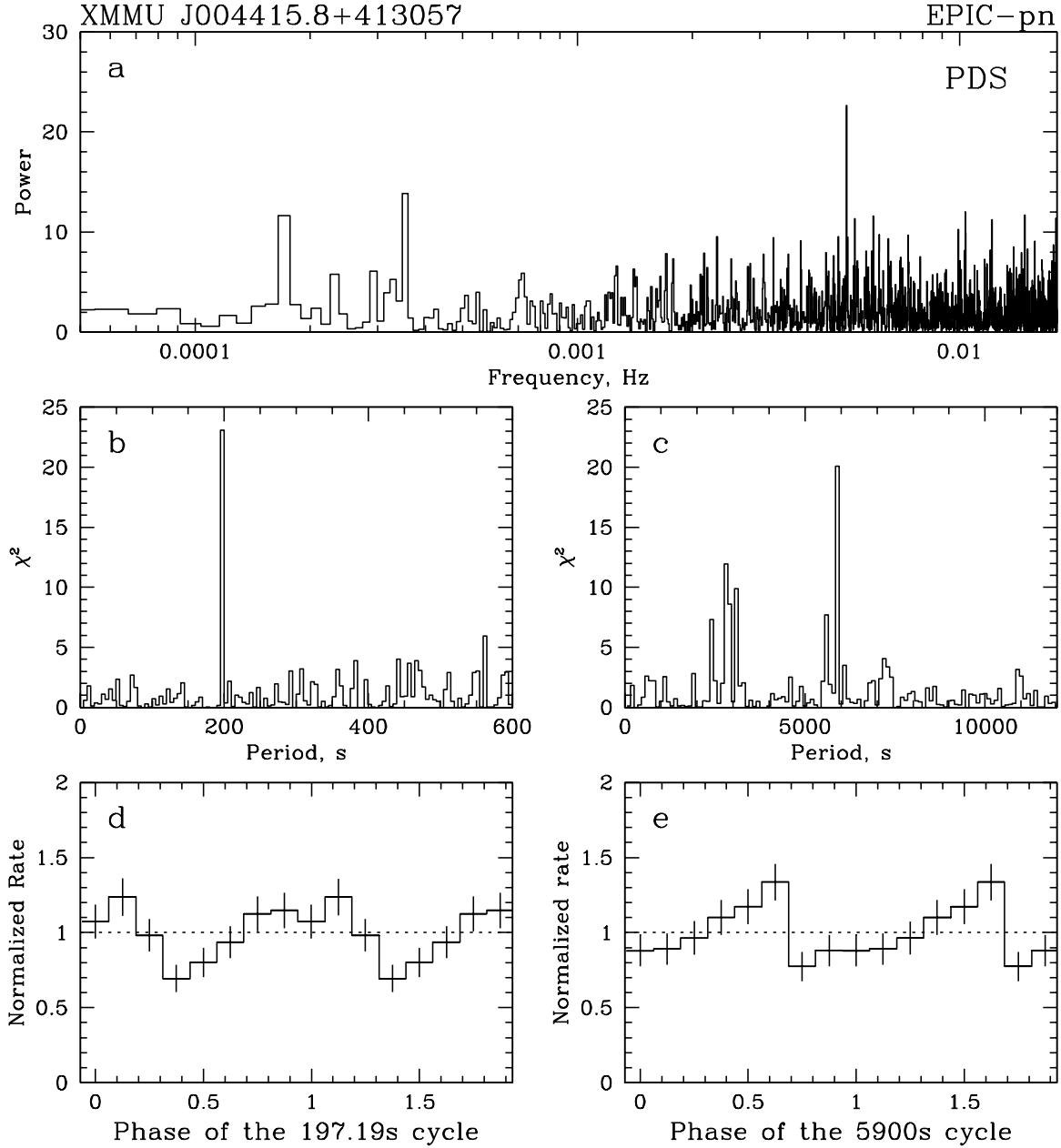}
\caption{\small The results of timing analysis of the light curve of XMMU J004415.8+413057 
(EPIC-pn data, $0.3 - 7.0$ keV energy band). {\em (a)} Power density spectrum of 
XMMU J004415.8+413057 showing possible X-ray pulsations with a period of 197 s and 
a slower 5900 s modulation. {\em (b,c)} Resulting $\chi^{2}$ distributions for the 
epoch-folding analysis of the X-ray light curve of the source. The best-fit periods 
of the modulations are 197.19 s and 5900 s respectively. {\em (d,e)} Normalized light 
curves of the source, folded at the best-fit periods of 197.19 s and 5900 s. 
\label{TR_087_timing_1}}
\end{figure}

The analysis of the timing properties of XMMU J004415.8+413057 revealed probable coherent 
pulsations of the X-ray flux with a period of 197.19(5) s (Fig. \ref{TR_087_timing_1}{\em a,b}). 
We estimated statistical significance of the corresponding peak in the source PDS. The probability 
that any one of the 1300 frequency bins in the PDS would have a noise value exceeding the power 
level of this peak is $\sim 1 \%$ (\cite{Vaughan94}). Figure \ref{TR_087_timing_1}{\em d} 
shows EPIC-pn light curve of XMMU J004415.8+413057 folded on the 197.19 s best period of the 
pulsation. The pulse profile shows a nearly sinusoidal form with $\sim 40 \%$ flux modulation. 

Another feature of the source PDS is a group of frequency peaks corresponding to the $\sim 5900$ 
s slow modulation (Fig. \ref{TR_087_timing_1}{\em a,c}). Figure \ref{TR_087_timing_1}{\em e} 
shows EPIC-pn light curve of XMMU J004415.8+413057 folded on the 5900 s period of this modulation. 

\begin{figure}
\epsfxsize=18cm
\epsffile{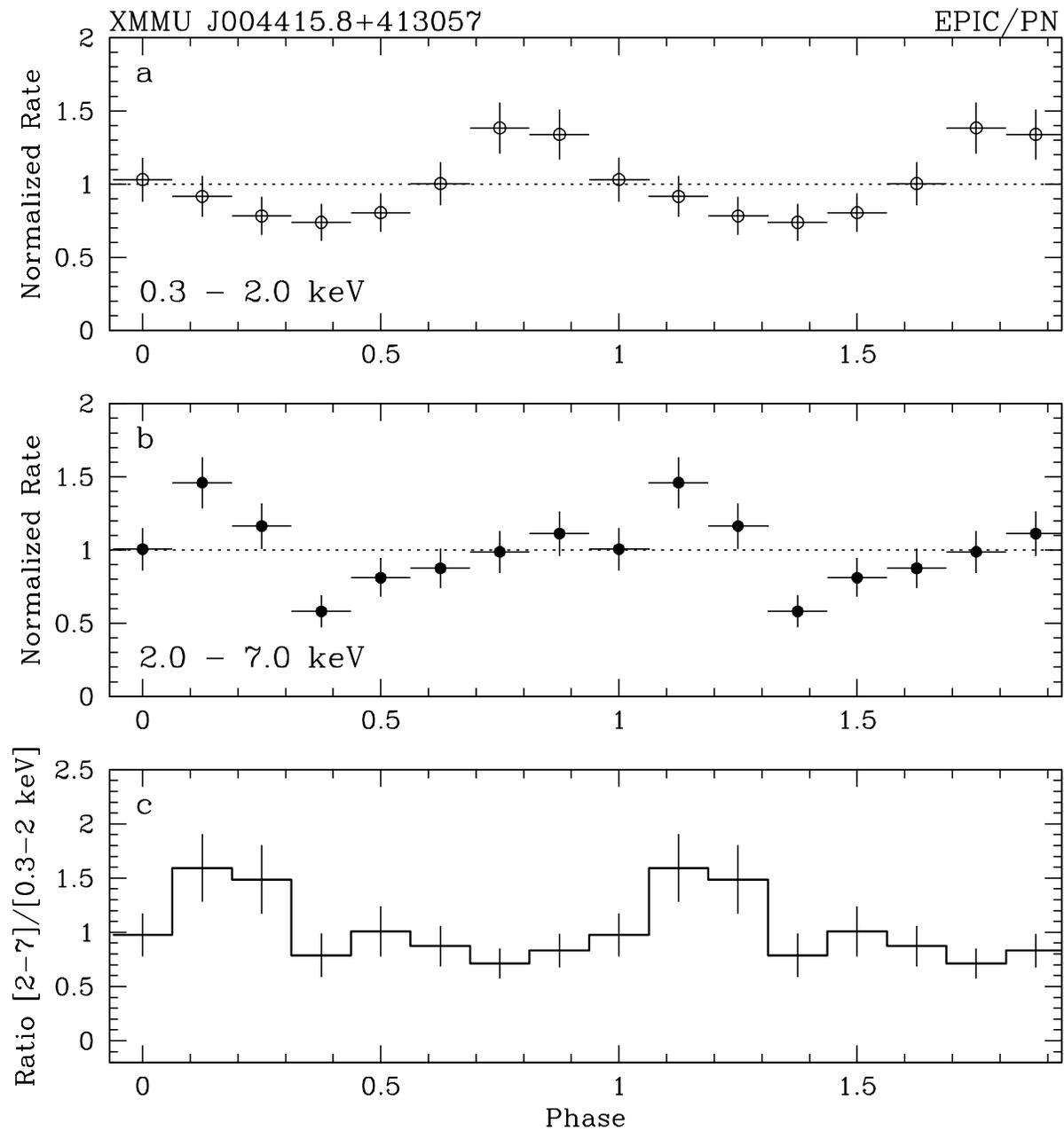}
\caption{\small {\em (a,b)} X-ray light curves of XMMU J004415.8+413057 in the $0.3 - 2.0$ and 
$2.0 - 7.0$ keV energy bands taken with EPIC-pn, folded on a period of 197.19 s. {\em (c)} 
Hardness of the source spectrum as a function of phase of 197.19 s modulation. 
\label{TR_087_timing_2}}
\end{figure}

In order to investigate the energy dependence of the 197-s pulsations, we constructed light curves 
in the $0.3 - 2.0$ and $2.0 - 7.0$ keV energy bands folded on the best period 
(Fig. \ref{TR_087_timing_2}). The energy-resolved pulse profiles show marginal differences between 
the soft and hard energies with excess of the hard emission at phase $\sim 0.15$. 

Together with the X-ray spectrum, probable 197-s pulsations strongly support the identification of 
XMMU J004415.8+413057 as a first accreting pulsar candidate in M31 and possible HMXB. The extreme 
level of low energy absorption in the spectrum (at least 10 times Galactic foreground value) could 
be consistent with absorption within HMXB system or source location in the region with high density 
of the neutral gas (e.g. molecular cloud). The $5900$-s modulation of the X-ray flux could be 
manifestation of some instability time scale of the accretion flow in this system. A significant 
variability of the the source on a time scales of several years could imply possible 
transient/recurrent nature of the source. 

\subsection{Supersoft X-ray transient XMMU J004414.1+412206}
The supersoft X-ray transient source XMMU J004414.1+412206 was discovered on 2002 January 5 using {\em XMM} 
(\cite{xmm_circ02}). The {\em XMM} discovery was confirmed by detection of the source with {\em Chandra} 
during Jan. 8 and 16, 2002 observations (\cite{chandra_circ02}). The search for the optical counterparts 
did not yield any object brighter than $m_{v} \sim 20$ within {\em XMM} and {\em Chandra} source error boxes.
The X-ray luminosity of XMMU J004414.1+412206 declined from $\sim 8 \times 10^{36}$ ergs s$^{-1}$ on Jan. 5 
to $\sim 5 \times 10^{36}$ ergs s$^{-1}$ on Jan. 8 (\cite{chandra_circ02}) for an assumed distance of 760 kpc. 
The energy spectra of the source obtained with {\em XMM-Newton}/EPIC-MOS1 and MOS2 detectors on January 5 
are shown in Figure \ref{spec_TR_fig}. The approximation of the EPIC-MOS spectra with the absorbed blackbody 
radiation model gives a characteristic temperature of $\sim 34$ eV and a relatively high absorbing column of 
$\sim 4 \times 10^{21}$ cm$^{-2}$ (Table \ref{spec_par_TR}; Figure \ref{spec_TR_fig}). 

The energy spectrum of XMMU J004414.1+412206 is typical for a supersoft X-ray sources, AM Her systems, 
isolated white dwarfs and X-ray dim neutron stars. Its luminosity of $\sim 8 \times 10^{36}$ ergs s$^{-1}$ 
for a distance of 760 kpc is too high to originate from an AM Her system in M31 (\cite{Ramsay94}). On the 
other hand, the transient behavior of the object poses a serious problem for a foreground isolated white 
dwarf or a neutron star identification. The absence of the bright optical objects in the error box of 
XMMU J004414.1+412206 excludes an explanation as an AM Her system located within our Galaxy.

The remaining possibility is that XMMU J004414.1+412206 is similar to the Galactic supersoft sources. Then 
its X-ray emission may be interpreted as a result of thermonuclear burning of the accreted matter on the 
surface of the white dwarf (see \cite{KVDH97} for a review). The transient behavior of the source hints 
that it may be a classical Nova in the supersoft X-ray spectral phase, several tens of days after the peak 
of the outburst (Kahabka \& van den Heuvel 1997). The X-ray source XMMU J004414.1+412206 is the fourth 
bright supersoft transient detected in M31.

\section{GLOBULAR CLUSTER CANDIDATES}
The positions of six bright X-ray sources are consistent with globular cluster candidates in M31 
(Table \ref{source_ID}). We performed a detailed study of their spectral and temporal properties using 
the data of {\em XMM-Newton} observations. 

\subsection{Spectral properties}
\begin{figure}
\epsfxsize=18cm
\epsffile{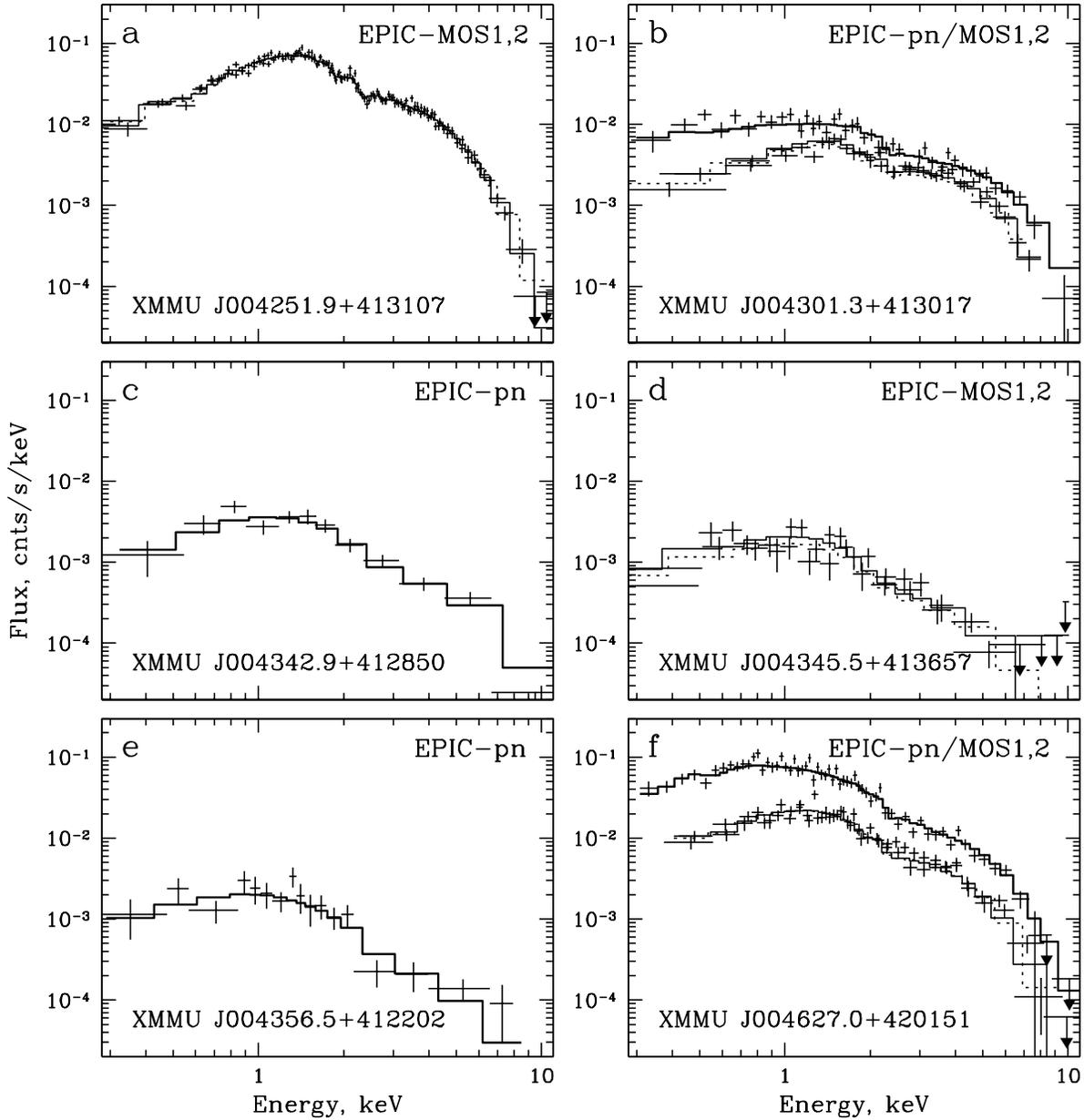}
\caption{\small {\em XMM-Newton}/EPIC spectra of the bright X-ray sources coincident with 
globular cluster candidates (EPIC-pn is always the upper spectrum and the two MOS spectra 
overlap). Corresponding model fits to the EPIC-pn, MOS1 and MOS2 spectra (Table \ref{spec_par_GCS}) 
are shown with thick, thin and dotted histograms respectively. {\em Panel a}: the X-ray 
source XMMU J004251.9+413107 ($\# 1$), EPIC-MOS1 and MOS2 data (obs. $\# 1$), fit by absorbed 
Comptonization model. {\em Panel b}: the X-ray source XMMU J004301.3+413017 ($\# 2$), 
EPIC-pn ($\# 4$), MOS1 and MOS2 data (obs. $\# 1$), fit by absorbed Comptonization 
model. {\em Panel c}: the X-ray source XMMU J004342.9+412850 ($\# 4$), 
EPIC-pn data (obs. $\# 1$), fit by absorbed simple power law model. {\em Panel d}: 
the X-ray source XMMU J004345.5+413657 ($\# 5$), EPIC-MOS1 and MOS2 data (obs. $\# 1$), 
fit by absorbed simple power law model. {\em Panel e}: the X-ray source XMMU 
J004356.5+412202 ($\# 7$), EPIC-pn data (obs. $\# 4$), fit by absorbed simple power 
law model. {\em Panel f}: the X-ray source XMMU J004627.0+420151 ($\# 27$), EPIC-pn, 
MOS1 and MOS2 data (obs. $\# 2$), fit by absorbed Comptonization model. 
\label{spec_GCS_fig}}
\end{figure}

The spectra of globular cluster candidates were fitted with a variety of spectral models using XSPEC v11. 
The results of fitting these models to the source spectra are given in Table \ref{spec_par_GCS}. The 
spectra of all these objects are relatively hard (Fig. \ref{spec_GCS_fig}) and can be generally described 
by an absorbed simple power law model with photon index of $\sim 0.8 - 1.9$ and an equivalent absorbing 
column of $\sim (0.4 - 2.6) \times 10^{21}$ cm$^{-2}$ (Table \ref{spec_par_GCS}). The corresponding 
isotropic luminosities of the XMM globular cluster sources differ by almost two orders of magnitude and 
fall between $3.7 \times 10^{36}$ and $3.3 \times 10^{38}$ ergs s$^{-1}$ in the $0.3 - 10$ keV energy band, 
assuming a distance of 760 kpc. The Galactic hydrogen column in the direction of M31 is about 
$\sim 7 \times 10^{20}$ cm$^{-2}$ (\cite{DL90}); thus within measurement errors our results are consistent 
with additional absorption due to the interstellar matter in M31 and within the system. 

For the three bright sources with luminosities above $\sim 5\times 10^{37}$ ergs s$^{-1}$, 
XMMU J004251.9+413107, XMMU J004301.4+413017 and XMMU J004627.0+420151 models with quasi-exponential cut-off 
at $\sim 4 - 8$ keV describe the energy spectra significantly better than a simple power law 
(Table \ref{spec_par_GCS}; Fig. \ref{spec_GCS_fig}). We used a power law with exponential cut-off and 
Comptonization models to approximate the spectra of these sources (Table \ref{spec_par_GCS}). 

For the Comptonization model approximation, we used the XSPEC model COMPTT (\cite{ST80,T94,TL95}). This 
model includes a self-consistent calculation of the spectrum produced by the Comptonization of the soft photons 
in a hot plasma. It contains as free parameters the temperature of the Comptonizing electrons, $kT_{e}$, the 
plasma optical depth with respect to the electron scattering, $\tau$ and the temperature of the input Wien soft 
photon distribution, $kT_{0}$. A spherical geometry was assumed for the Comptonizing region. 

The spectra of many luminous Galactic LMXB are well fit with a two component model consisting of a black 
body-like component which might represent emission from an optically thick accretion disk or from the neutron 
star surface, together with a Comptonized component which may be interpreted as emission from a hot inner disk 
region or a boundary layer between the disk and a neutron star. We used such a two-component model to approximate 
spectra of the brightest globular cluster source, XMMU J004251.9+413107. For the soft component we used 
disk-blackbody component (\cite{Mitsuda84}). This model has two parameters, the effective radius, 
$r_{in} \sqrt{cos i}$, where $r_{in}$ is the inner radius of the disk, $i$ is the inclination angle of the disk 
and $k T_{in}$ is the maximum effective temperature in the disk. For the Comptonized component the XSPEC model 
COMPTT described above was used. The best-fit parameters of the model are shown in Table 
\ref{spec_par_GCS_two_comp}.

The spectral properties of the bright GC X-ray sources detected in the {\em XMM} observations of the northern 
disk regions of M31 are similar to that of the bright Galactic low mass X-ray binaries located in the bulge 
and globular clusters (\cite{Sidoli01}; \cite{Iaria01}; \cite{DiSalvo01}). Two out of total six GC X-ray sources 
in our sample were found to have X-ray luminosities exceeding $10^{38}$ ergs s$^{-1}$, making them brighter 
than any known Galactic GC X-ray source. 

\subsection{Variability}
The comparisons based on the broad-band spectral properties are not sufficient to establish a neutron star 
nature or rule out a black hole nature for the M31 globular cluster sources. On the other hand, the study of 
their short-term variability can provide a definitive answer, if Type I X-ray bursts or X-ray pulsations are 
observed. We searched for both types of variability in the M31 GC data. Unfortunately, the observed source count 
rates for most sources are too low to probe timescales shorter than $\sim 20 - 30$ seconds. We did not find 
evidence of X-ray pulsations or short X-ray bursts in {\em XMM-Newton} data. 

\subsection{Correlation between X-ray and optical properties}
We used the results of optical observations of M31 GC candidates (\cite{Huchra91}; 
\cite{Barmby00}; \cite{Perrett02}) to study how the properties of the X-ray sources depend on the overall 
properties of the globular clusters hosting them.  

\begin{figure}
\epsfxsize=18cm
\epsffile{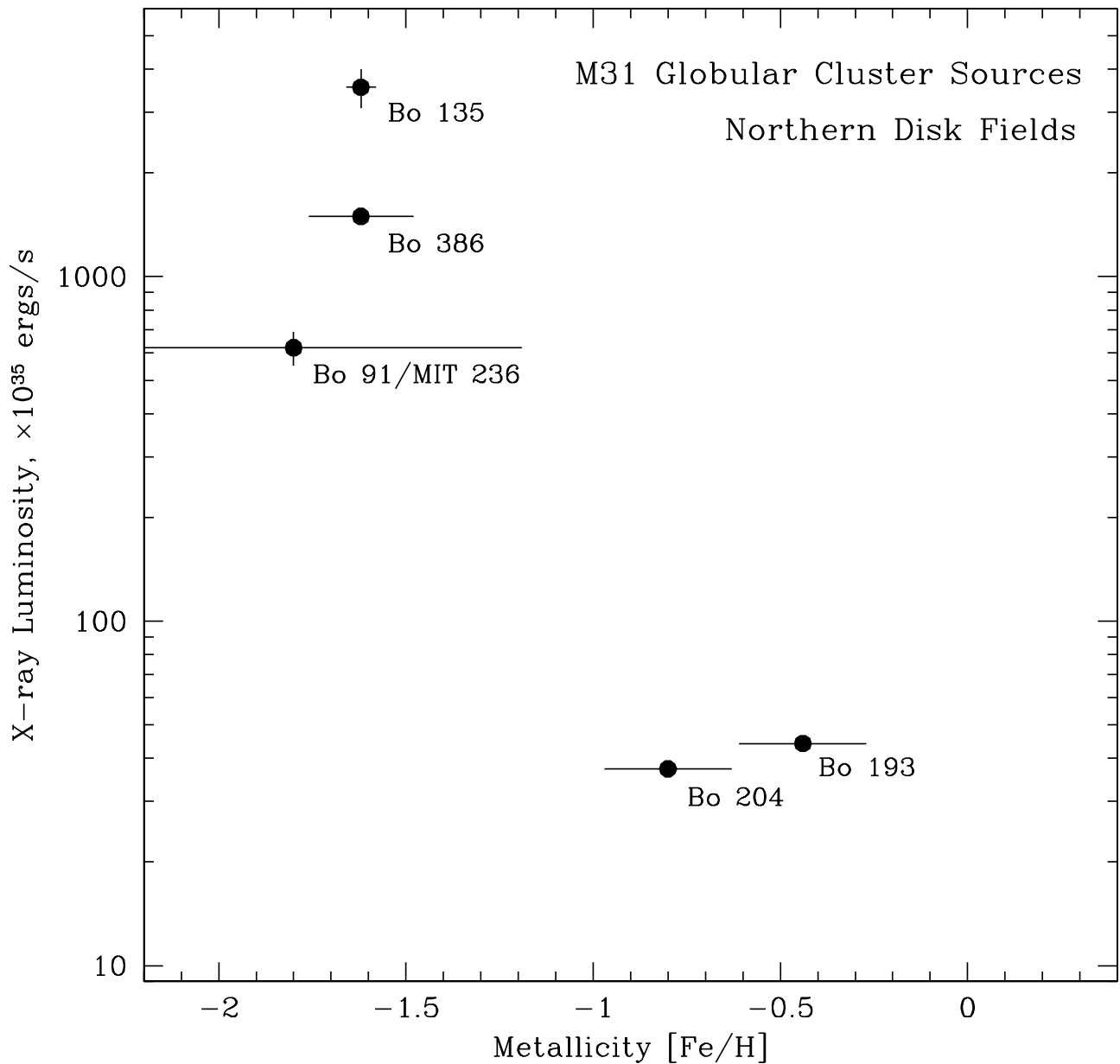}
\caption{\small The absorbed $0.3 - 10$ keV X-ray luminosity of GC sources detected in the 
{\em XMM-Newton} observations of the northern disk of M31 vs. metallicity of their host globular 
clusters. The optical data are from Huchra, Brodie $\&$ Kent (1991), Barmby et al. (2000) and 
Perrett et al. (2002). \label{GC_met_lum}}
\end{figure}

The Galactic globular clusters hosting bright LMXBs were found to be both denser and more 
metal-rich (\cite{Sidoli01}). We studied the effect of globular cluster metallicity on the 
luminosity of the M31 GC sources from our sample. The resulting dependence is shown in 
Fig. \ref{GC_met_lum} ({\em left panel}). The tendency of the brightest sources to reside 
in the more metal poor clusters ([Fe/H]$< -1.5$) is obvious. This behavior seems to be in 
general disagreement with possible correlation between the metallicity and X-ray luminosity 
found for the bright Galactic GC sources (\cite{Sidoli01}). The limited statistics of our sample 
(5 M31 GC X-ray sources) does not allow us to draw conclusions on the applicability of our 
results to the whole M31 globular cluster X-ray source population. 

\section{SUPERNOVA REMNANTS}
There are three SNR candidates in the lists by Braun \& Walterbos (1993) and Magnier et 
al. (1995) coincident with bright XMM sources (sources $\#\#$ 3, 17 and 21 in Table 
\ref{source_ID}). All three sources were previously detected with {\em ROSAT}/PSPC and 
HRI and identified as SNR (Supper et al. 1997, 2001; \cite{Magnier97}). 

\begin{figure}
\epsfxsize=18cm
\epsffile{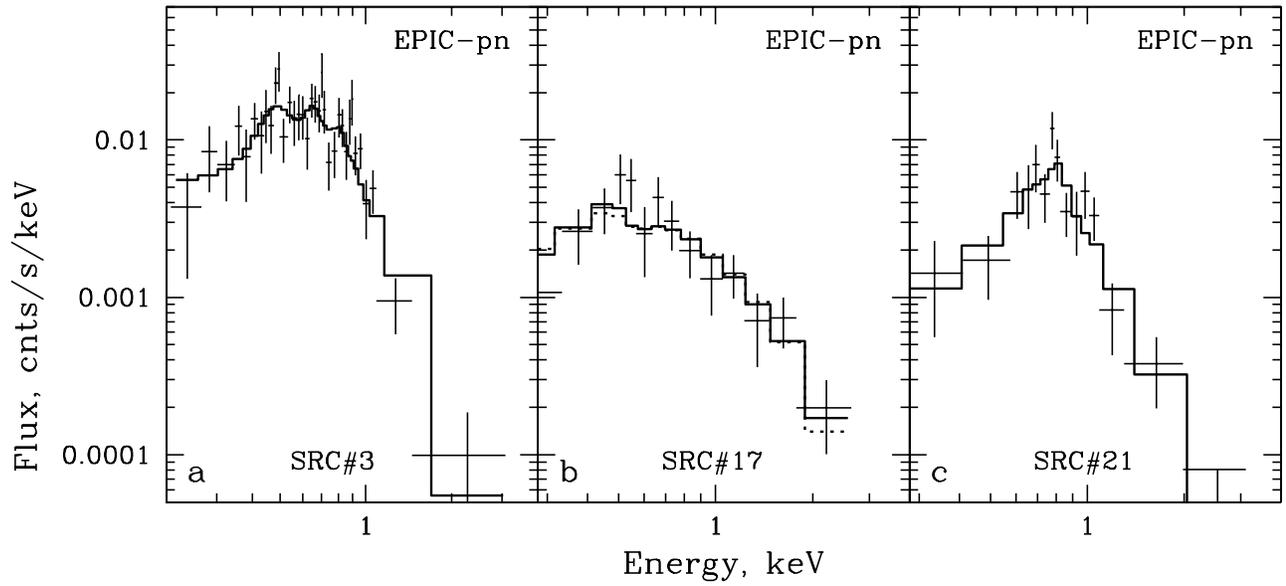}
\caption{\small EPIC-pn count spectra of three bright SNR candidates: XMMU J004339.1+412654 
({\em a}), XMMU J004451.1+412907 ({\em b}) and XMMU J004513.9+413614 ({\em c}). For 
each source the best-fit absorbed Raymond-Smith thermal plasma models (RS) are shown 
with solid histograms. The approximation of the spectrum of XMMU J004451.1+412907 with 
absorbed simple power law is shown in the {\em panel b} with dotted histogram. 
\label{spec_SNR_fig}}
\end{figure}

The energy spectra of these 3 SNR candidates are shown in Fig. \ref{spec_SNR_fig}. 
We fitted the spectra of SNR candidates with various single component spectral models 
including a simple power law, thermal bremsstrahlung, black body, Raymond-Smith thermal 
plasma (RS) (\cite{RS}), and non-equilibrium ionization collisional plasma (NEI) models 
with interstellar absorption. The results of the analytical approximation of the 
{\em XMM}/EPIC-pn data for SNR candidates are shown in Table \ref{spec_par_SNR}. 

RS and NEI models are often used to study X-ray emission from supernova remnants. These 
models give the best approximation to the data for two SNR candidate sources $\#\#$ 3 and 
21, which show clear presence of the emission lines in their spectra (Fig. \ref{spec_SNR_fig}; 
Table \ref{spec_par_SNR}). 

{\em XMMU J004339.1+412654 ($\# 3$)} Except for the RS and NEI models, other simple 
spectral models do not provide acceptable fits to the data ($\chi^{2}$/$\nu > 2$). For 
the RS and NEI models, we first fixed the abundances at solar value (\cite{AG89}). This 
set of parameters, however, left bump-like residuals indicating 
that we had overestimated the contribution from O-K, Ne-K and Fe-L shell emission lines. 
Then we fixed the abundances based on the values determined from optical spectroscopy of 
the optical counterpart to XMMU J004339.1+412654 (BW57 in Table \ref{source_ID}) (\cite{Blair82}): 
[N/H]=$1.1\times10^{-4}$, [O/H]=$3.7\times10^{-4}$, [S/H]=$9.3\times10^{-6}$ and 
[Ne/H]=$1.2\times10^{-4}$, corresponding to 0.98, 0.44, 0.57 and 1.00 solar abundance. The 
Fe abundance was fixed at solar value. Finally, for the RS model we allowed the O, Ne and Fe 
abundances to be free parameters. The fit improved slightly. The best-fit values of the 
parameters for the approximation of XMMU J004339.1+412654 spectrum with the RS and NEI models 
are listed in Table \ref{spec_par_SNR}. Depending on the type of model and a set of model 
parameters (Table \ref{spec_par_SNR}), the estimated corresponding absorbed luminosity of 
XMMU J004339.1+412654 in the $0.3 - 3.0$ keV energy band lies in the range of 
$(1.8 - 2.0)\times 10^{36}$ ergs s$^{-1}$. 

{\em XMMU J004451.1+412907 ($\# 17$)} Due to the limited statistics, both RS and simple 
power law models provide satisfactory approximation to the spectrum of the fainter SNR 
candidate source XMMU J004451.1+412907 (Fig. \ref{spec_SNR_fig}). Although the EPIC-pn 
spectrum of XMMU J004451.1+412907 can be fitted well with RS model with free elemental 
abundances, we were unable to obtain meaningful constraints on the abundances. Therefore, 
we assume element abundances to be proportional to the solar values, reducing the number of 
free abundance parameters to one. The best-fit spectral parameters for the approximation of 
the spectrum of XMMU J004451.1+412907 with RS model are listed in Table \ref{spec_par_SNR}. 
The corresponding X-ray luminosity in the $0.3 - 3.0$ keV energy band is 
$\sim 8 \times 10^{35}$ ergs s$^{-1}$. The absorbed power law model with photon index of 
$3.4^{+0.7}_{-0.4}$, an equivalent hydrogen column of $1.6^{+1.0}_{-0.4} \times 10^{21}$ 
cm$^{-2}$ and $0.3 - 3.0$ keV flux of $1.3 \times 10^{-14}$ ergs s$^{-1}$ cm$^{-2}$ also 
gives satisfactory approximation to the data ($\chi^{2}/{\nu} = 0.6$) (Fig. \ref{spec_SNR_fig}). 

{\em XMMU J004513.9+413614 ($\# 21$)} We used RS model with interstellar absorption to 
approximate the spectrum of SNR candidate XMMU J004513.9+413614. We first fixed the 
element abundances at solar value (\cite{AG89}). Then we fixed the abundances based on 
the values determined from optical spectroscopy of the optical counterpart to 
XMMU J004339.1+412655 (BW39 in Table \ref{source_ID}) (\cite{Blair82}): 
[N/H]=$6.8\times10^{-5}$, [O/H]=$4.5\times10^{-4}$, [S/H]=$6.8\times10^{-6}$ and 
[Ne/H]=$1.2\times10^{-4}$, corresponding to 0.61, 0.53, 0.42 and 1.00 solar abundance, 
with the Fe abundance fixed at the solar value. Finally, we allowed the Fe abundance to 
vary freely. This fit, optimized with Fe at 0.18 solar, was by far the best (Fig. 
\ref{spec_SNR_fig}, {\em panel c}). The fit parameters are listed in Table \ref{spec_par_SNR}. 
The estimated $0.3 - 3.0$ keV X-ray luminosity of XMMU J004513.9+413614 lies in the 
range of $(0.8 - 1.1) \times 10^{36}$ ergs s$^{-1}$.

\section{FOREGROUND STARS}
\begin{figure}
\epsfxsize=18cm
\epsffile{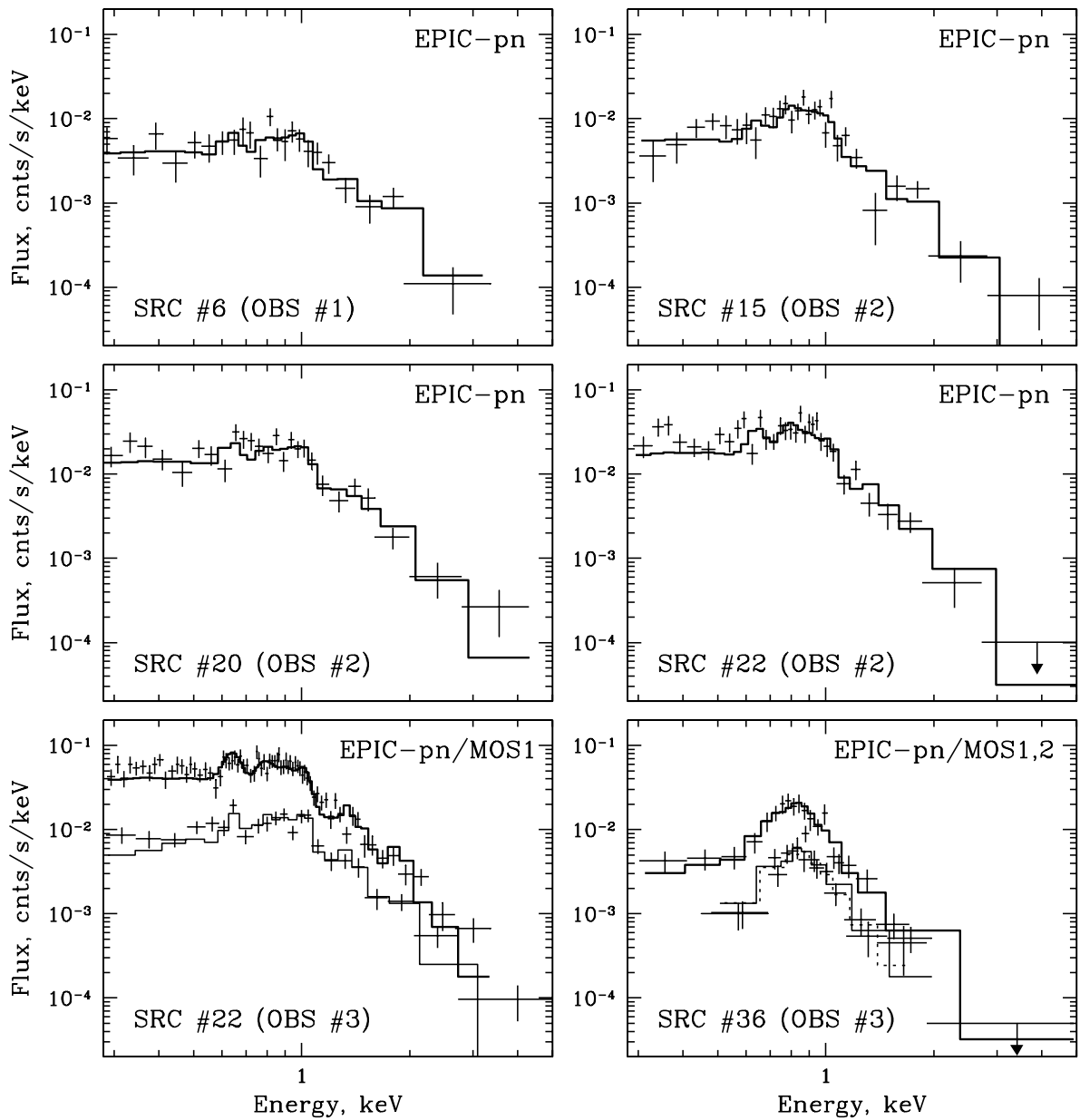}
\caption{\small EPIC count spectra of Galactic foreground star candidates. The absorbed plasma 
emission model (RS) fits are shown with thick (EPIC-pn), thin (EPIC-MOS1) and thin dotted 
(EPIC-MOS2) histograms (pn is always the upper spectrum and the two MOS spectra overlap). 
\label{spec_FRGS_fig}}
\end{figure}

We found 5 bright {\em XMM} sources to coincide with objects in the Magnier et al. 
(1992) and 2MASS catalogs of stars (Table \ref{source_ID}) with X-ray spectral 
properties consistent with that of Galactic foreground stars (Supper et al. 
1997; \cite{HDP03}). All these sources have soft X-ray spectra without 
significant emission above $2.5$ keV and occupy a distinctive area on the X-ray 
color-color diagram (Fig. \ref{colors}). The energy spectra of bright sources 
identified with Galactic foreground stars are shown in Figure \ref{spec_FRGS_fig}. 
To approximate their energy spectra, we used the model of the emission from the 
optically thin thermal plasma (XSPEC Raymond-Smith model, RS) with characteristic 
temperature between 0.6 and 0.8 keV and relatively low iron abundance of $0.1 - 0.6$ 
solar, corrected for the low energy absorption (Table \ref{spec_par_FRGS}). For all 
objects the absorbing column is well below the Galactic value in the direction of M31 
(Table \ref{spec_par_FRGS}), further supporting identification as foreground stars. The 
X-ray spectra and optical properties of 3 of these sources suggest emission from late-type 
stars with active coronae, also consistent with colors of their optical counterparts. 

\begin{figure}
\epsfxsize=18cm
\epsffile{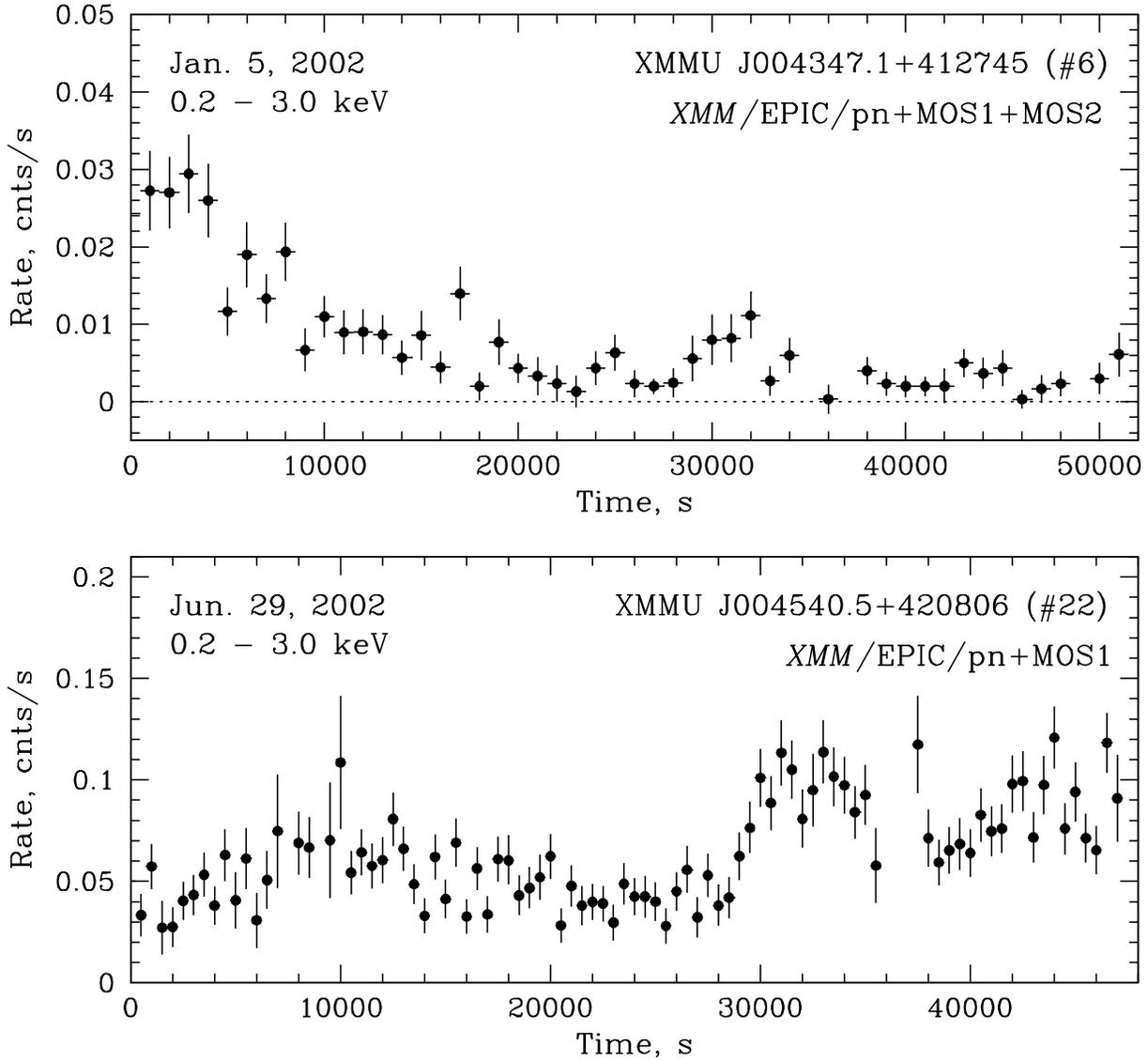}
\caption{\small {\em Upper panel:} X-ray light curve of foreground star candidate source 
XMMU J004347.1+412745 (src $\#6$ in Table \ref{source_ID}) during 2002 January 5 
{\em XMM-Newton} observation (Obs. $\# 1$), obtained from combined data of EPIC-pn, 
MOS1 and MOS2 cameras, the $0.2 - 3.0$ keV energy range, and a 1000 s time resolution. 
{\em Lower panel:} X-ray light curve of foreground star candidate source 
XMMU J004540.5+420806 (src $\#22$ in Table \ref{source_ID}) during 2002 June 29 
{\em XMM-Newton} observation (Obs. $\# 3$), obtained from combined data of EPIC-pn and  
MOS1 cameras the $0.2 - 3.0$ keV energy range, and a 500 s time resolution. 
\label{lc_FRGS_fig}}
\end{figure}

Two sources, XMMU J004347.1+412745 ($\#6$) and XMMU J004540.5+420806 ($\#22$) demonstrate 
significant variability during 2002 Jan. 5 and Jun. 29 {\em XMM} observations. The X-ray 
lightcurves of these sources in the $0.2 - 3.0$ keV energy range are shown in Fig. 
\ref{lc_FRGS_fig}. 

The source XMMU J004347.1+412745 shows a dramatic change of the X-ray flux during 2002, 
Jan. 5 observation (obs. $\# 1$). The source flux history can be described as a combination 
of a relatively smooth decline (source flux drops $\sim 15$ times in the first 23 ks of 
observation) and a number of irregular flares (Fig. \ref{lc_FRGS_fig}). 

The source XMMU J004540.5+420806 shows significant variability of X-ray flux during 2002, 
Jun. 29 {\em XMM} observation (obs. $\# 3$). The source brightness changes with a peak flux 
level $\sim 4$ times higher than the low flux level (Fig. \ref{lc_FRGS_fig}). It should be 
noted, that during the 2002, Jan. 26 {\em XMM} observation (obs. $\# 2$) the X-ray flux of 
XMMU J004540.5+420806 appears to be remarkably stable.

\section{RADIO SOURCES}
\begin{figure}
\epsfxsize=18cm
\epsffile{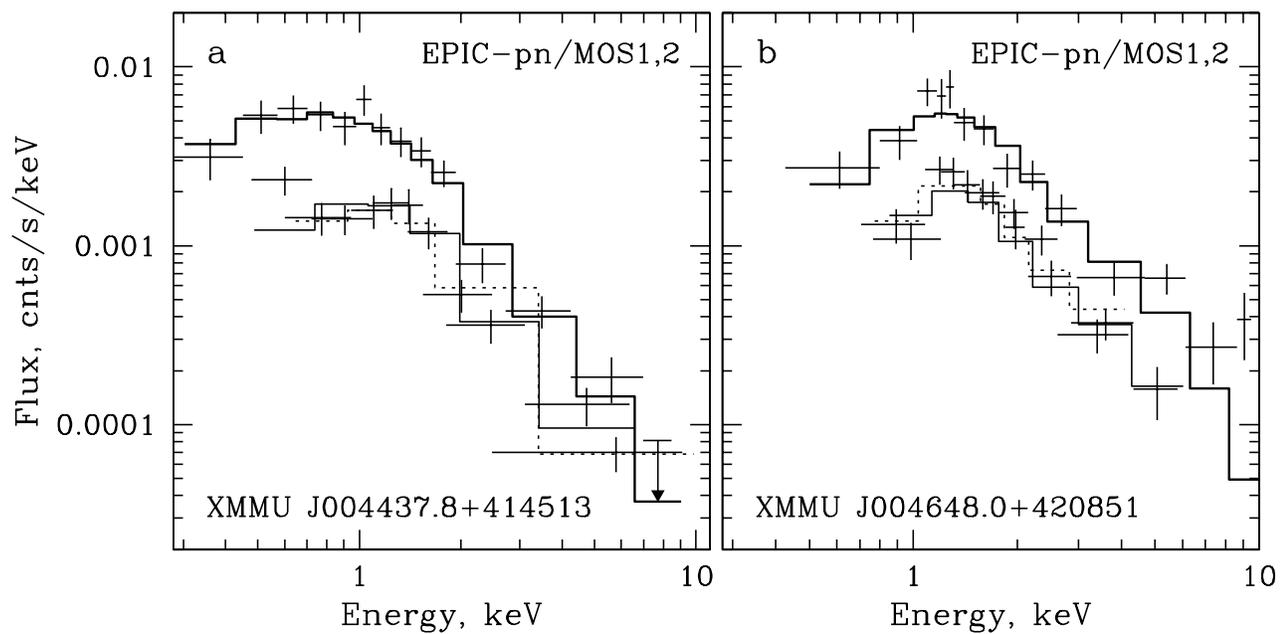}
\caption{\small EPIC count spectra of two X-ray sources coincident with bright radio sources: 
XMMU J004437.8+414513 ({\em a}) and XMMU J004648.0+420851 ({\em b}). For each source 
the absorbed power law model best fit to the three EPIC spectra are shown with thick 
(EPIC-pn), thin (EPIC-MOS1) and thin dotted (EPIC-MOS2) histograms (pn is always the 
upper spectrum and the two MOS spectra overlap). \label{spec_RADIO_fig}}
\end{figure}

The positions of the two bright {\em XMM} sources not identified as SNR candidates 
($\#\# 16, 28$ in Table \ref{source_ID}) coincide with bright radio sources detected with 
VLA all-sky (\cite{nvss}) and earlier surveys. These two objects are most likely radio 
galaxies in the background of M31. Both sources have X-ray spectra presented by 
absorbed power laws with indices between 1.8 and 2.2, typical for this source class (Fig. 
\ref{spec_RADIO_fig}; Table \ref{spec_par_radio}). 

\section{CLUSTER OF GALAXIES XMMU J004624.8+420420} 
One of the sources detected in the Obs. $\#\# 2$ and $3$, XMMU J004624.8+420420 
($\# 26$ in Table \ref{source_ID}) was classified as distant cluster of galaxies 
based on its spatial extent and X-ray spectrum (\cite{kotov03}). X-ray emission of 
the cluster is clearly detectable up to the radial distance of $3\arcmin$ from its 
center. X-ray emission from the source was detected previously by ROSAT, and 
cataloged as RX J0046.4+4204 (\cite{Supper01}), although it was not recognized as 
an extended object and a galaxy cluster. A joint spectral fit to the data of 
EPIC-MOS2 and pn cameras with the Raymond-Smith thermal model gives a temperature 
of a hot gas inside the cluster, $kT_{gas} = 4.3 \pm 0.3$ keV. The 
X-ray spectrum of XMMU J004624.8+420420 shows prominent iron emission line indicating 
cluster redshift of $z = 0.293$. For a cosmological model with $H_0$ = 50 km s$^{-1}$ 
Mpc$^{-1}$, $\Omega_M$ = 0.3 and $\Omega_{\Lambda}$ = 0.7 we derive a bolometric 
luminosity of 1.6 $\times$ $10^{45}$ erg/s (\cite{kotov03}). We examined optical 
images of the cluster region and found a concentration of highly reddened galaxies 
coincident with the central part of XMMU J004624.8+420420, further supporting our 
interpretation. In addition, we found a relatively bright radio source 
NVSS J004625+420406 (\cite{nvss}) located only $16\arcsec$ away from the emission 
center of XMMU J004624.8+420420, indicating that it might be radio galaxy belonging 
to the cluster.   

\section{POSSIBLE AGN AND OTHER UNIDENTIFIED SOURCES}
The majority of the remaining 18 X-ray sources detected in our survey consists of AGN 
in the background of M31 or X-ray binaries belonging to the galaxy. For some of these 
sources there is a close stellar-like counterpart (sources $\#\# 31,37$), indicating 
possible AGN origin. For the source $\# 33$ a background galaxy is likely optical 
counterpart. The spectra of most of these objects are hard and can be represented by 
power laws with indices between 1.5 and 2.3 (Table \ref{spec_par_other}). The 
corresponding values of the interstellar absorption inferred from their spectra are 
close to or well above expected Galactic value (Table \ref{spec_par_other}). For 
some of these sources a high value of measured column density combined with very low 
brightness of the possible optical counterpart could be explained by AGN intrinsic 
obscuration. More reliable optical identifications of these obscured AGN candidates 
are needed to confirm or reject their AGN nature.  
  
\section{X-RAY COLOR-COLOR DIAGRAMS}
In order to facilitate comparison between spectral properties of different source 
classes detected in our observations of the northern disk of M31, we constructed 
their X-ray color-color diagram. We calculated the total number of counts for each 
source using its corrected EPIC-pn spectra in three energy bands: the soft band 
($0.3 - 1.0$ keV), medium band ($1.0 - 2.0$ keV) and hard band ($2.0 - 7.0$ keV). 
Two X-ray colors were defined for each source as: $HR1 = (S - M)/T$ (soft color) and 
$HR2 = (H - M)/T$ (hard color), where $S, M,$ and $H$ are the counts in soft, medium 
and hard bands respectively, and $T$ is the total number of source counts in the 
$0.3 - 7.0$ keV energy range.

\begin{figure}
\epsfxsize=18cm
\epsffile{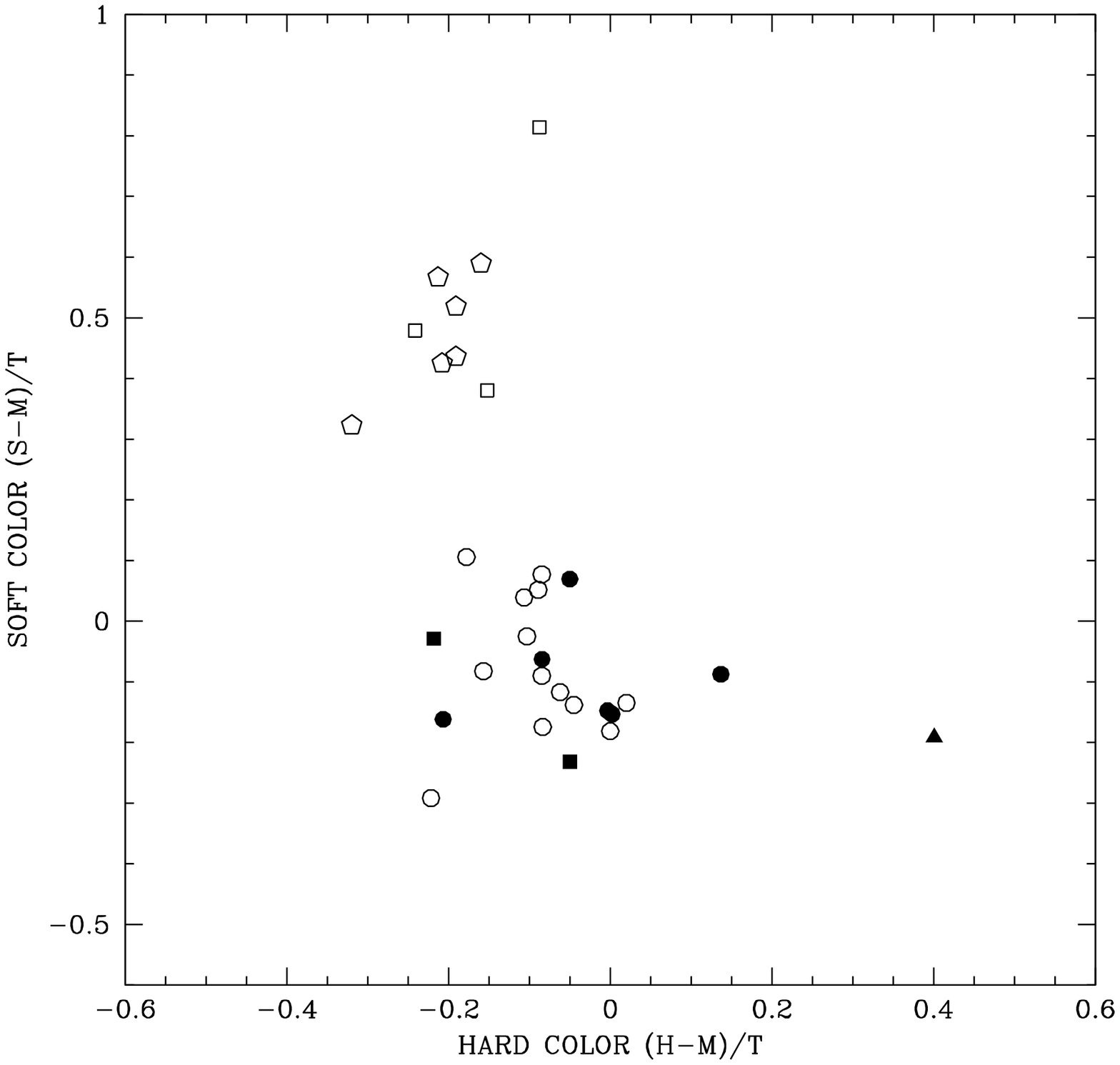}
\caption{\small X-ray color-color diagram for the bright sources detected with {\em XMM}/EPIC-pn. 
The energy bands are: the soft band ($0.3 - 1.0$ keV), medium band ($1.0 - 2.0$ keV) and 
hard band ($2.0 - 7.0$ keV). Two X-ray colors were defined for each source as: 
$HR1 = (S - M)/T$ (soft color) and $HR2 = (H - M)/T$ (hard color), where $S, M,$ and $H$ are 
the counts in soft, medium and hard bands respectively, and $T$ is the total number of source 
counts in the $0.3 - 7.0$ keV energy range. SNR candidates, foreground stars, globular cluster 
candidates, HMXB pulsar candidate, unidentified sources and radio sources are shown with 
{\em open squares}, {\em open stars}, {\em filled circles}, {\em filled triangle}, {\em open circles}, 
and {\em filled squares}. 
\label{colors}}
\end{figure}

Figure \ref{colors} shows the X-ray color-color diagram for bright X-ray sources 
detected in the {\em XMM-Newton}/EPIC-pn observations of the northern disk of M31. 
There are two distinct concentrations of sources in this diagram. The first group 
with $HR1 > 0.3$ includes intrinsically soft sources -- SNR candidates and foreground 
stars. The other densely populated group includes harder sources: GCS, radio sources 
and unidentified sources (presumably background AGN and X-ray binaries in M31). The 
X-ray pulsar candidate (possible HMXB) demonstrates harder and much more absorbed 
spectrum, and lies at an extreme position with $HR1 \sim -0.2$ and $HR2 \sim 0.4$.

Fig. \ref{colors} demonstrates that in most cases it is difficult to establish 
source type using X-ray color alone. Although it helps to outline the difference between 
sources with soft and hard spectra (like thermal supernova remnants and stars vs. typical 
high-mass X-ray binaries or black hole candidates in a hard state), it shows no difference 
between other physically different classes like, for example, X-ray binaries and Crab-like 
supernova remnants. The combination of low-energy absorption and limited instrument bandpass 
have significant effect on the source position on the color-color diagram (\cite{DiStefano02}; 
\cite{Prestwich03}). For example, the source with intrinsically super-soft spectrum, 
if highly absorbed, can easily ``migrate'' to the region on the color diagram normally 
occupied by sources with much harder spectra (\cite{DiStefano02}). Therefore, additional 
information, such as X-ray variability, luminosity and source counterparts at other wavelengths, 
is needed to classify the X-ray source.  

\section{DIFFUSE X-RAY EMISSION FROM THE DISK OF M31}
Previous X-ray studies of M31 were mostly concentrated on the central region of the galaxy. 
They revealed the presence of the unresolved soft X-ray emission component in the bulge of 
M31 distinct from the emission of point-like sources (\cite{Primini93}; \cite{Shirey01}). 
The situation with the outer disk of M31 is less clear. Previous observations were not 
sensitive enough or lacked spatial resolution to fully address a question of the existence 
of the diffuse X-ray component in the disk. Using the data of {\em Einstein} observations, 
Trinchieri $\&$ Fabbiano (1991) found no requirement for a diffuse component after exclusion 
of the detected point-like sources. Based on the later observations with {\em ROSAT}/PSPC, 
West, Barber $\&$ Folgheraiter (1997) report on a detection of a residual unresolved X-ray 
emission from the outer disk of M31. They conclude that the bulk of this emission could be 
attributed to the normal stars in the disk of M31.

\begin{figure}
\epsfxsize=19cm
\epsffile{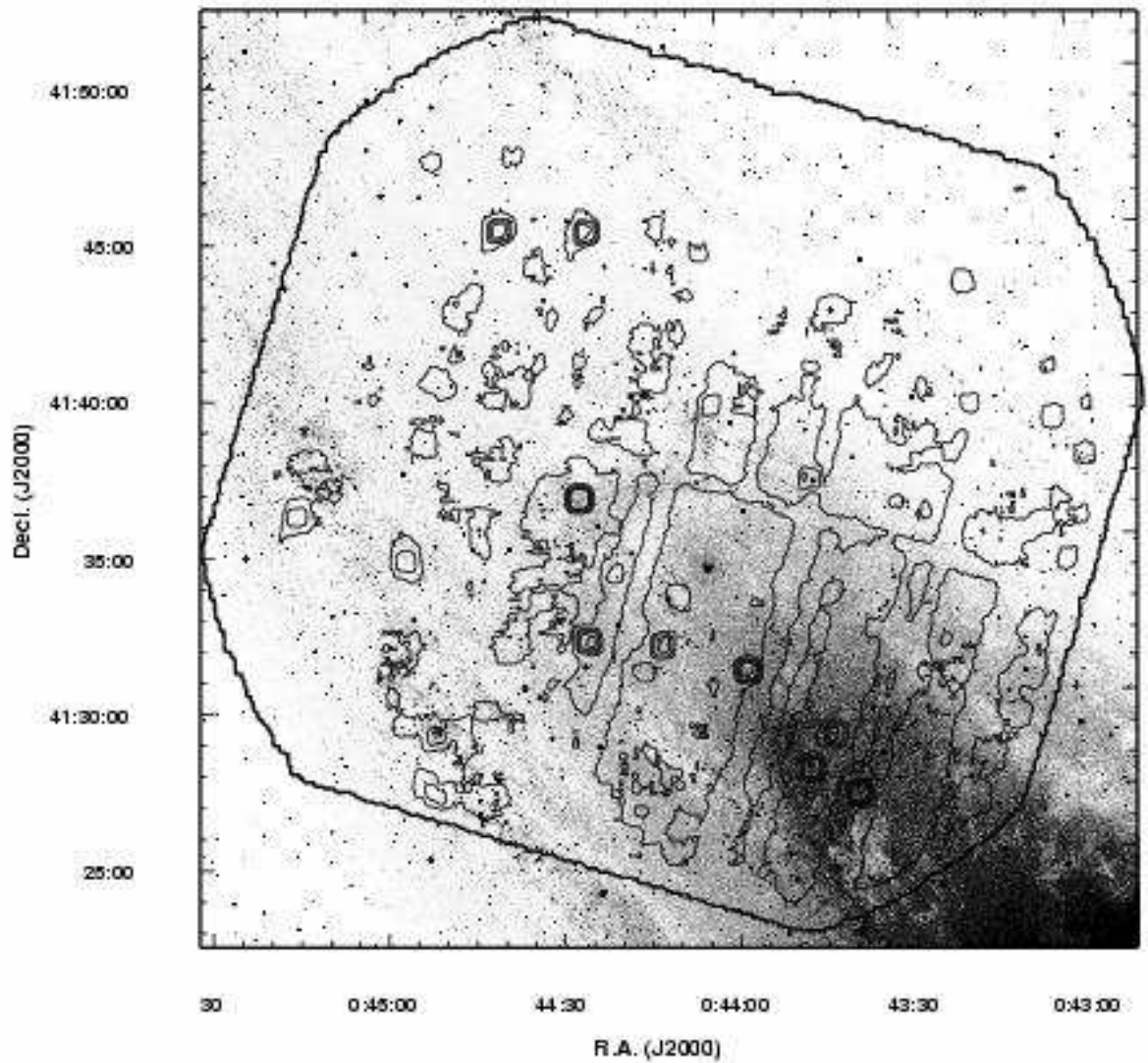}
\caption{\small Optical image of the northern disk of M31 from the Second Generation 
Digitized Sky Survey blue plate with {\em XMM}/EPIC-pn X-ray contours overlayed. X-ray 
contours were produced using the EPIC-pn image from the Obs. $\#1$ in the $0.2 - 1.5$ 
keV energy band, convolved with circularly symmetric Gaussian function with 
$\sigma = 20\arcsec$. The outer thick contour marks the instrument FOV. The contour 
levels start at $3.3\times 10^{-3}$ cnts s$^{-1}$ arcmin$^{-2}$ spaced by equal intervals 
of $3.3\times 10^{-3}$ cnts s$^{-1}$ arcmin$^{-2}$. X-ray contours show extended unresolved 
X-ray emission tracing the structure of the optical disk as well as numerous point-like 
sources. The cross-like pattern of gaps in contours is due to the gaps between EPIC-pn CCDs 
and excluded bad detector columns. 
\label{diffuse_pn_overlay}}
\end{figure}

The observation of M31 North 1 field revealed the presence of soft unresolved soft 
X-ray emission from the disk of M31. The unresolved emission follows the pattern of the spiral 
structure and can be traced up to the distance of $\sim 0.5\arcdeg$ ($\sim 7$ kpc at 760 
kpc) from the center of the galaxy (Fig. \ref{diffuse_pn_overlay}). The unresolved emission 
includes a large region along the major axis of M31 and a number of smaller regions in the 
spiral arms. The unresolved emission covers the ``inner star-forming ring'' of M31 
(\cite{Haas98}) and has several local enhancements that are clearly associated with regions 
of recent star formation (\cite{schmidtobreick00}).
 
\begin{figure}
\epsfxsize=18cm
\epsffile{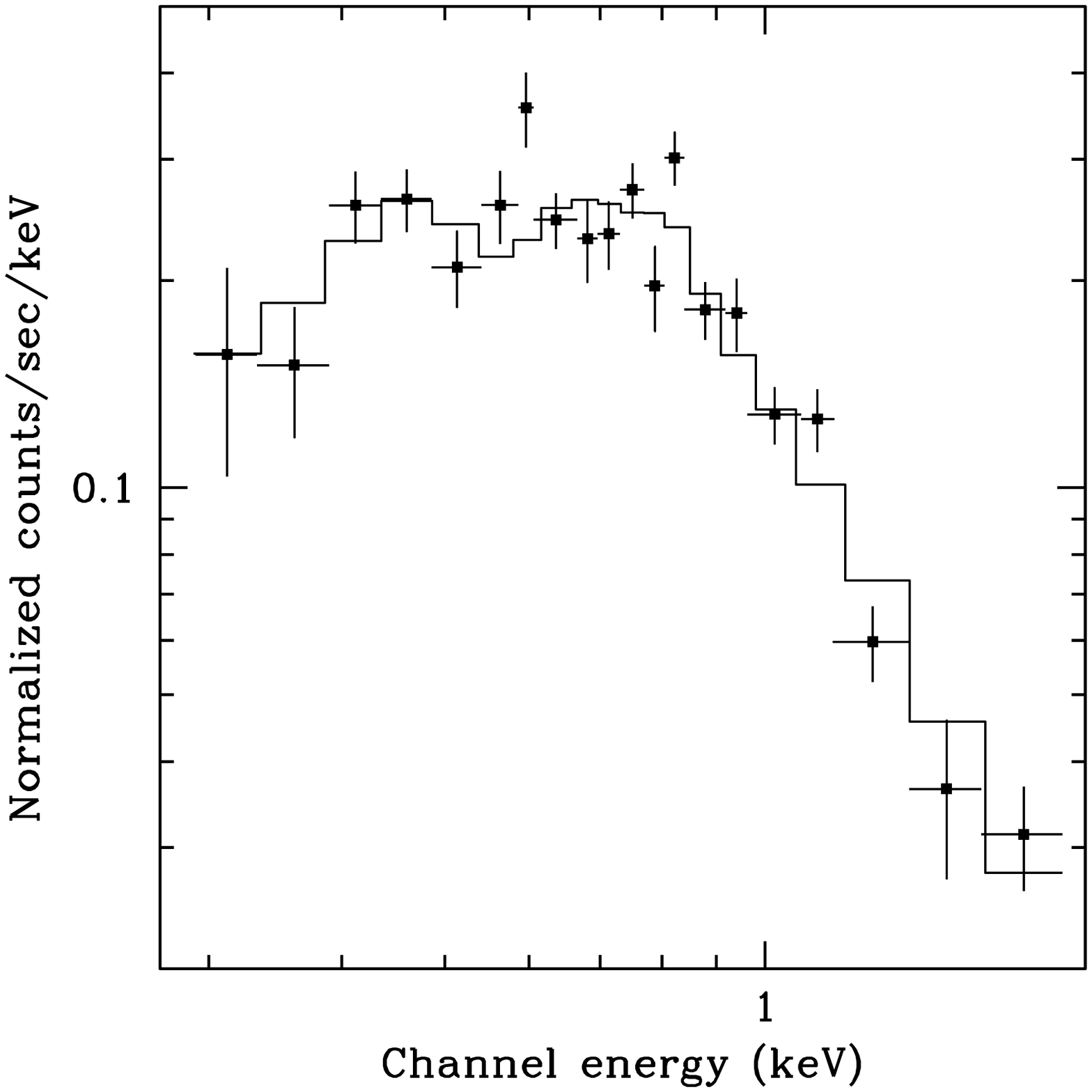}
\caption{\small Representative EPIC-pn count spectrum of the unresolved X-ray emission from 
the disk of M31. The approximation with thermal plasma model (RS) is shown with thin 
histogram. \label{diffuse_spec_pn}}
\end{figure}

To characterize the diffuse emission from the disk of M31, we extracted the EPIC-pn spectrum 
from $\sim 65$ arcmin$^{2}$ region in the disk at the projected distance of $\sim 5$ kpc from 
the center, using the background region outside the disk and eliminating point-like sources 
from both the source and background regions. To reduce contamination from faint SNR, regions 
containing undetected optical SNR candidates (Braun \& Walterbos 1993; Magnier et al. 1995) 
were also excluded from source and background extraction regions. The resulting normalized 
count spectrum is shown in Fig. \ref{diffuse_spec_pn}. The spectrum of the unresolved emission 
is soft and shows possible presence of spectral lines. A crude approximation of its shape in 
the $0.3 - 2.0$ keV energy range with a power law model gives a value of photon index of 
$\sim 4.6$. The spectrum of unresolved emission can be adequately approximated by two-component 
spectral model including the Raymond-Smith optically thin thermal plasma emission model with 
a temperature of $0.26\pm0.05$ keV and normalized $0.3 - 2.0$ keV flux of 
$(3\pm0.3)\times 10^{-15}$ ergs s$^{-1}$ cm$^{-2}$ arcmin$^{-2}$ and a weak power law 
component with photon index of $\sim 1.7$ contributing $\lesssim 25\%$ of the total 
luminosity in the $0.3 - 2.0$ keV energy band (Fig. \ref{diffuse_spec_pn}). 

The spectrum of unresolved emission differs significantly from the cumulative spectrum of 
faint point-like sources in the same region, which can be presented by absorbed power law 
model of photon index $1.8\pm0.1$ and absorbing column density $N_{\rm H} \sim 10^{21}$. 
The spectral properties of the unresolved emission are somewhat similar to that of the 
thermal SNR detected in our observations (see Section 6), but both selection procedure and 
great spatial extent of the unresolved emission make serious spectral contamination from 
SNR unlikely. This argues strongly that the bulk of the diffuse X-ray emission in the disk 
of M31 is not simply due to the faint, unresolved X-ray binaries and SNR. It suggests that 
there is a significant hot interstellar gas contribution to the diffuse emission, analogous 
to the hot gas components found in the center of M31 (\cite{Shirey01}), in the disk of M33 
(\cite{Long96}; \cite{Pietsch03}), and in our own Galaxy. Another significant contributor 
to the diffuse emission component could be X-ray emission from normal stars in the disk of 
M31 (\cite{West97}).

A weak power law component in the spectrum of unresolved emission still can be due to the 
combined effect of faint X-ray binaries below our detection limit and some remaining 
contribution from extracted point-like sources.

\section{SUMMARY}
A series of deep {\em XMM-Newton} observations of northern disk of M31 allow a detailed 
study of spectral and temporal properties of detected X-ray sources down to flux levels 
of $\lesssim 10^{-13}$ ergs s$^{-1}$ cm$^{-2}$. The unprecedented sensitivity of EPIC 
cameras allowed us to detect and study the unresolved soft X-ray emission associated with 
disk of M31. 

We present the results of X-ray analysis of thirty seven of the brightest sources detected 
in the M31 northern disk fields. Combining the results of X-ray analysis with available data 
at other wavelengths, we were able to classify $\sim 50 \%$, or 19 out of 37 sources. 

Two sources in our sample were previously unknown: the hard X-ray source XMMU J004415.8+413057 
and a transient supersoft source XMMU J004414.1+412206. A detailed analysis of the timing 
properties of XMMU J004415.8+413057 revealed possible X-ray pulsations with a period of 197 s. 
The combination of the unique spectral and timing properties of XMMU J004415.8+413057 would 
make it the first accreting X-ray pulsar and HMXB candidate detected in M31. 

Six X-ray sources were identified with M31 globular clusters (GC). The spectral properties of 
GC sources were found to be strikingly similar to that of the bright Galactic low mass X-ray 
binaries located in the bulge and globular clusters (\cite{Sidoli01}; \cite{Iaria01}; 
\cite{DiSalvo01}). Two GC X-ray sources (XMMU J004251.9+413107 and XMMU J004627.0+420151) have 
estimated persistent isotropic luminosities above $10^{38}$ ergs s$^{-1}$, making them brighter 
than any known Galactic GC X-ray source. The comparison of the X-ray properties of GC sources 
with optical properties of their host globular clusters has shown that brighter sources tend to 
reside in more metal poor clusters (Fig \ref{GC_met_lum}, {\em left panel}). 

Three X-ray sources were identified with SNR based on their X-ray spectra and positional 
coincidence with SNR candidates from optical and radio surveys. We found five bright X-ray 
sources to coincide with Galactic foreground stars. Two of them demonstrate a remarkable 
variability of X-ray flux on a time scale of individual {\em XMM-Newton} observations. 
Two X-ray sources coincide with radio sources and are probably distant radio galaxies/AGN. 

The properties of the remaining 18 bright X-ray sources detected in our survey are consistent 
with AGN in the background of M31 and X-ray binaries belonging to M31. Many of them show high 
values of low-energy absorption, which combined with extreme faintness of their possible 
optical counterparts makes them good candidates for an obscured AGN. 

The {\em XMM-Newton} observations of M31 North 1 field revealed the presence of soft 
unresolved X-ray emission from the disk of M31. It follows the pattern of the spiral arms and 
can be traced up to the distance of $\sim 0.5\arcdeg$ ($\sim 7$ kpc at 760 kpc) from the center 
of the galaxy. The spectral properties and spatial distribution of the unresolved emission 
suggest that there is a significant hot interstellar gas contribution to the diffuse emission, 
analogous to the hot gas components found in the center of M31, in the disks of other external 
galaxies and in our own Galaxy.

\section{Acknowledgments}
Part of this work was done during a summer workshop at the Aspen Center for Physics, S. T. and 
W. P. are grateful to the Center for their hospitality. Support for this work was provided 
through NASA Grant NAG5-12390. XMM-Newton is an ESA Science Mission with instruments and 
contributions directly funded by ESA Member states and the USA (NASA). This research has made 
use of data obtained through the High Energy Astrophysics Science Archive Research Center Online 
Service, provided by the NASA/Goddard Space Flight Center. This publication makes use of data 
products from the Two Micron All Sky Survey, which is a joint project of the University 
of Massachussetts and the Infrared Processing and Analysis Center/California Institute of 
Technology, funded by NASA and NSF.

\begin{table}
\small
\caption{XMM-Newton observations of M31 North1, North2 and North3 fields. 
\label{obslog}}
\small
\begin{tabular}{ccccccccc}
\hline
\hline
Obs. $\#$ &Date, UT & $T_{\rm start}$, UT & Field & Obs. ID  & RA (J2000)$^{a}$ & 
Dec (J2000)$^{a}$ & Exp.(MOS)$^{b}$ & Exp.(pn)$^{b}$\\
& &(h:m:s)&&&  (h:m:s)   &(d:m:s)&(ks)&(ks)\\             
\hline
$\#1$&2002 Jan 05 &06:28:31&M31 North1&0109270701&00:44:01.0&41:35:57.0&57.3&54.7\\
$\#2$&2002 Jan 26 &16:51:03&M31 North2&0109270301&00:45:20.0&41:56:09.0&29.1&25.3\\
$\#3$&2002 Jun 29 &08:59:02&M31 North3&0109270401&00:46:38.0&42:16:20.0&51.5&36.5\\
$\#4$&2002 Jan 06 &18:07:17&M31 Core&0112570101&00:42:43.0&41:15:46.1&63.0&61.0\\ 
\hline
\end{tabular}
\begin{list}{}{}
\item[$^{a}$] -- coordinates of the center of the field of view
\item[$^{b}$] -- instrument exposure used in the analysis 
\end{list}
\end{table}

\begin{table}
\small
\caption{Sample of selected bright X-ray sources in the EPIC observations of northern disk 
fields of M31 \label{source_ID}}
\small
\begin{tabular}{cclcccl}
\hline
\hline
Source& {\em XMM} Source Name &Optical/IR/Radio &Offset$^{b}$& SHP97$^{c}$ &Offset$^{d}$& Comment\\ 
  ID  &     (XMMU ...)        &    ID$^{a}$     & ($\arcsec$)&             &($\arcsec$) &        \\
\hline
 1& J004251.9+413107 & Bo 135                  & 1.4 & 205     & 1.8 & GCS \\
 2& J004301.3+413017 & Bo 91/MIT 236           & 0.9 & 218     & 5.3 & GCS \\
 3& J004339.1+412654 & BW57/MA 3-069           & 2.3/6.9 & 249 & 5.6 & SNR \\
  &                  & W85 180                 & 2.2 &         &     & \\
 4& J004342.9+412850 & MIT 311                 & 1.3 & 250     & 1.4 & GCS \\ 
  &                  & MLV92 315558            & 1.0 &         &     & \\
 5& J004345.5+413657 & Bo 193                  & 1.2 & 253     & 4.4 & GCS \\
 6& J004347.1+412745 & 2MASS 004347.18+412744.5& 1.6 &         &     & Foreground star   \\
  &                  & MLV92 308951            & 1.8 &         &     &                   \\
 7& J004356.5+412202 & Bo204                   & 0.6 & 261     & 3.5 & GCS \\
 8& J004357.5+413054 & MLV92 328135            & 1.3 & 268     & 2.1 & \\
 9& J004402.7+413927 & MLV92 379667	           & 1.0 & 266     & 4.1 & \\
10& J004412.0+413147 & MLV92 333307            & 0.9 & 272     & 4.7 & \\
11& J004414.1+412206 &                         &     &         &     & Supersoft transient\\
12& J004415.8+413057 &                         &     &         &     & Transient, HMXB? 197s pulsar\\
13& J004422.5+414507 & MLV92 394966            & 0.3 & 278     & 3.3 & \\
14& J004424.7+413200 & MLV92 334722            & 0.7 & 279     & 12  & \\
  &                  & MLV92 334827            & 1.7 &         &     & \\
15& J004425.5+413633 & 2MASS 004425.57+413633.4& 0.6 & 281     & 6.0 & Foreground star\\                
  &                  & MLV92 362312            & 0.4 &         &     & \\
16& J004437.8+414513 & NVSS J004437+414511     & 2.0 &         &     & Radio source\\
  &                  & W85 207                 & 6.5 &         &     & \\
17& J004451.1+412907 & BW 33                   & 2.6 & 288     & 16  & SNR\\
18& J004452.2+412717 & MLV92 306564            & 2.2 & 289     & 2.4 & \\
  &                  & MLV92 306271            & 2.5 &         &     & \\   
19& J004455.4+413440 &                         &     &         &     & \\
20& J004456.3+415936 & 2MASS 004456.37+415936.8& 0.4 & 292     & 7.0 & Foreground star\\  
  &                  & MLV92 436498            & 0.6 &         &     & \\
21& J004513.9+413614 & BW 39/MA 2-049          & 1.7/4.1 & 297 & 9.1 & SNR\\
  &                  & W85 231                 & 4.7 &         &     & \\
22& J004540.5+420806 & 2MASS 004540.54+420806.8& 0.8 & 316     & 4.0 & Foreground star \\   
  &                  & MLV92 450979            & 0.6 &         &     & \\
23& J004544.9+415858 &                         &     & 317     & 2.6 & \\
24& J004611.5+420826 &                         &     & 332     & 5.3 & \\
25& J004619.9+421441 & MLV92 462752            & 1.6 &         &     & \\ 
  &                  & MLV92 462656            & 4.3 &         &     & \\
26& J004624.8+420420 & NVSS J004625+420406     &  16 & 348     & 19  & Cluster of Galaxies at $z=0.3$\\    
27& J004627.0+420151 & Bo 386                  & 1.2 & 349     & 4.0 & GCS \\
28& J004648.0+420851 & NVSS J004648+420855     & 3.8 & 353     & 1.8 & Radio source\\
  &                  & W85 252                 & 2.8 &         &     & \\
29& J004651.8+421951 &                         &     & 354     & 12  & \\
30& J004651.9+421504 &                         &     &         &     & \\
31& J004655.4+422049 & 2MASS 004655.51+422050.1& 1.6 & 355     & 4.1 & \\
  &                  & MLV92 472612            & 1.4 &         &     & \\
32& J004703.6+420449 & MLV92 444881            & 3.7 & 357     & 5.6 & \\                        
33& J004706.5+422211 & MLV92 474281            & 1.9 & 359     & 12  & \\  
34& J004713.3+422045 & MLV92 472554            & 2.2 & 361     & 11  & \\
35& J004725.2+422118 & MLV92 473224            & 2.0 &         &     & \\
36& J004726.1+422157 & 2MASS 004726.12+422158.4& 1.6 & 369     & 6.1 & Foreground star\\ 
  &                  & MLV92 474007            & 1.4 &         &     & \\
37& J004748.3+421932 & MLV92 470983            & 1.8 & 384     & 7.1 & \\
\hline
\end{tabular}

\begin{list}{}
\item $^{a}$ -- source identifications beginning with BW refer to 
the SNR candidates listed in Braun $\&$ Walterbos (1993). Identifications 
beginning with MA refer to the SNR candidates from Magnier et al. (1995). 
Identifications beginning with Bo refer to M31 Globular Cluster candidates 
listed in Table IV of Battistini et al. (1987). Identifications beginning 
with MIT refer to M31 Globular cluster candidates from Magnier (1993).
\item $^{b}$ -- angular distance between {\em XMM} source position and the 
position of its possible optical counterpart.
\item $^{c}$ -- SHP97 refer to M31 {\em ROSAT}/PSPC catalogue entries from 
Supper et al. (1997, 2001).
\item $^{d}$ -- offset between {\em XMM} and {\em ROSAT} source positions.
\end{list}
\end{table}

\begin{table}
\small
\caption{Spectral fit results for bright sources discovered with {\em XMM-Newton}. 
\label{spec_par_TR}}
\small
\begin{tabular}{cc}
\hline
\hline
Parameter  & \\
\hline
                 & XMMU J004415.8+413057$^{a}$ ($\# 12$)\\
\hline
\multicolumn{2}{c}{Absorbed Power Law}\\
\hline
Photon Index                              & $1.43^{+0.19}_{-0.11}$ \\
$N_{\rm H}$, $\times 10^{22}$ cm$^{-2}$   & $1.36^{+0.47}_{-0.26}$ \\
Flux$^{b}$                                & $1.29\pm0.07$          \\
$\chi^{2}$(d.o.f.)                        & $16.1(20)$             \\
\hline
                 & XMMU J004414.1+412206$^{c}$ ($\# 11$)\\
\hline
\multicolumn{2}{c}{Absorbed Black Body Model}\\
\hline
$kT_{\rm bb}$, eV                         & $34\pm3$                \\
$N_{\rm H}$, $\times 10^{22}$ cm$^{-2}$   & $0.43\pm0.03$           \\
Flux$^{d}$                                & $1.11\pm0.04$           \\
$\chi^{2}$(dof)                           & $88.6(33)$              \\
\hline
\end{tabular}

\begin{list}{}{}
\item $^{a}$ -- EPIC-pn data, $0.3 - 10$ keV energy range
\item $^{b}$ -- Absorbed model flux in the $0.3 - 10.0$ keV energy range in 
units of $10^{-13}$ erg s$^{-1}$ cm$^{-2}$
\item $^{c}$ -- EPIC-MOS1 and MOS2 data, $0.2 - 1.0$ keV energy range
\item $^{d}$ -- Absorbed model flux in the $0.3 - 1.0$ keV energy range in 
units of $10^{-13}$ erg s$^{-1}$ cm$^{-2}$
\end{list}
\end{table}

\begin{table}
\small
\caption{Bright GCS spectral fit results ({\em XMM-Newton}/EPIC data, $0.3 - 10$ keV energy range). 
\label{spec_par_GCS}}
\small
\begin{tabular}{ccccccccc}
\hline
\hline
ID $^{a}$ & & & & & & & & Remarks\\
\hline
              & \multicolumn{7}{c}{Model: Absorbed Power Law (POWERLAW*WABS)}&        \\
\hline
              & Photon   & N$_{\rm H}^{b}$           &Flux$^{c}$&$\chi^{2}$  & &&$L_{X}^{d}$&     \\
              & Index    &                           &          & (dof)    & && &    \\
\hline
01&$1.57^{+0.03}_{-0.02}$&$0.26\pm0.01$&$52.95\pm0.42$&443.7(424)& &&$3651$&MOS1+MOS2\\
02&$0.83\pm0.04$&$0.07^{+0.01}_{-0.02}$&$8.91\pm0.19$&228.6(176)& &&$614$&MOS1+MOS2+pn\\
04&$1.61^{+0.14}_{-0.13}$&$0.23\pm0.05$&$0.75\pm0.05$&33.2(32)& &&$52$&pn\\
05&$1.61^{+0.11}_{-0.04}$&$0.04\pm0.03$&$0.64\pm0.03$&26.3(26)& &&$44$&MOS1+MOS2\\
07&$1.90\pm0.32$&$0.19\pm0.07$&$0.54\pm0.05$&14.6(16)& &&$37$&pn\\
27&$1.55^{+0.04}_{-0.03}$&$0.13\pm0.01$&$21.66\pm0.30$&270.2(237)& &&$1493$&MOS1+MOS2+pn\\
\hline
              & \multicolumn{7}{c}{Model: Absorbed Cutoff Power Law (CUTOFFPL*WABS)} & \\
\hline
              & Photon & Cutoff Energy & $N_{\rm H}^{b}$ & Flux$^{c}$ & $\chi^{2}$ & &$L_{X}^{d}$& \\
              & Index  &   (keV)       &                 &            & (dof)      & & & \\
\hline
01&$0.84\pm0.11$&$4.20^{+0.76}_{-0.57}$&$0.16^{+0.02}_{-0.01}$&$47.80\pm0.40$&$399.3(423)$&&$3296$&MOS1+MOS2\\
02&$0.18\pm0.09$&$4.86^{+1.12}_{-0.72}$&$<0.02$&$7.91\pm0.17$&$208.0(175)$&&$545$&MOS1+MOS2+pn\\
27&$1.16^{+0.16}_{-0.17}$&$7.49^{+5.33}_{-2.17}$&$0.09^{+0.01}_{-0.02}$&$20.44\pm0.29$&$264.4(236)$&&$1409$& MOS1+MOS2+pn\\ 
\hline
              & \multicolumn{7}{c}{Model: Absorbed Comptonization Model (COMPTT*WABS)} & \\
\hline
              & $kT_{0}$ & $kT_{e}$ & Optical Depth & $N_{\rm H}^{b}$& Flux$^{c}$ & $\chi^{2}$&$L_{X}^{d}$& \\
              &  (keV)   &  (keV)   &               &                &   & (dof) & & \\   
\hline
01&$0.37^{+0.01}_{-0.03}$&$1.88^{+0.23}_{-0.18}$&$19.5^{+0.9}_{-1.6}$&$<0.02$&$47.82\pm0.41$&
$398.7(422)$&$3297$&MOS1+MOS2\\
02&$0.03^{+0.07}_{-0.01}$&$1.76^{+0.14}_{-0.12}$&$34.6^{+2.8}_{-2.5}$&$0.07^{+0.01}_{-0.02}$&
$7.38\pm0.16$&$206.8(174)$&$509$&MOS1+MOS2+pn\\
27&$0.14\pm0.04$&$1.64^{+0.15}_{-0.16}$&$22.4^{+2.2}_{-2.3}$&$0.09\pm0.02$&$19.42\pm0.28$&$258.3(235)$&$1339$&MOS1+MOS2+pn\\
\hline 
\end{tabular}

\begin{list}{}{}
\item $^{a}$ -- Source number in Table \ref{source_ID}
\item $^{b}$ -- Equivalent hydrogen column depth in units of $10^{22}$ cm$^{-2}$ 
\item $^{c}$ -- Absorbed model flux in the $0.3 - 10$ keV energy range in 
units of $10^{-13}$ erg s$^{-1}$ cm$^{-2}$
\item $^{d}$ -- Absorbed isotropic source luminosity in the $0.3 - 10.0$ keV 
energy range in units of $10^{35}$ erg s$^{-1}$ assuming the distance of 760 kpc
\end{list}
\end{table}

\begin{table}
\small
\caption{Spectral fit results for the approximation of the EPIC-MOS spectra 
of XMMU J004251.9+413107 with combination of absorbed Comptonization and 
disk-blackbody models. \label{spec_par_GCS_two_comp}}
\small
\begin{tabular}{cc}
\hline
\hline
Parameter  & \\
\hline
                 & XMMU J004251.9+413107$^{a}$ ($\# 1$)\\
\hline
\multicolumn{2}{c}{(COMPTT+DISKBB)*WABS}\\
\hline
$N_{\rm H}$, $\times 10^{22}$ cm$^{-2}$   & $0.14\pm0.02$          \\
$kT_{0}$, keV                             & $0.58^{+0.09}_{-0.11}$ \\
$kT_{e}$, keV                             & $2.14^{+1.56}_{-0.43}$ \\
Optical Depth                             & $17.4^{+3.6}_{-11.1}$  \\
$kT_{in}$, keV                            & $0.53^{+0.13}_{-0.19}$ \\
$r_{in} \sqrt{cos i}$, km                 & $68^{+96}_{-30}$       \\
Flux$^{b}$                                & $48.54\pm0.41$         \\
$f_{soft}^{c}$                            & $0.23$                 \\
$\chi^{2}$(d.o.f.)                        & $391.9(420)$           \\
$L_{X}^{d}$                               & $3347$                 \\
\hline
\end{tabular}

\begin{list}{}{}
\item $^{a}$ -- EPIC-MOS1 and MOS2 data, $0.3 - 10$ keV energy range
\item $^{b}$ -- Absorbed model flux in the $0.3 - 10.0$ keV energy range in 
units of $10^{-13}$ erg s$^{-1}$ cm$^{-2}$
\item $^{c}$ -- Absorption-corrected ratio of the disk-blackbody flux to the 
total source flux in the $0.3 - 10.0$ keV energy band
\item $^{d}$ -- Absorbed isotropic source luminosity in the $0.3 - 10.0$ keV 
energy range in units of $10^{35}$ erg s$^{-1}$ assuming the distance of 760 kpc
\end{list}
\end{table}

\begin{table}
\small
\caption{Bright SNR candidate sources spectral fit results ({\em XMM-Newton}/EPIC data, absorbed RS and NEI models, 
{\em XMM}/EPIC-pn data $0.3 - 3.0$ keV energy range. \label{spec_par_SNR}}
\small
\begin{tabular}{ccccccccccc}
\hline
\hline
Model    & $kT_{\rm RS}/kT_{e}$&log$n_{e} t$&\multicolumn{5}{c}{Abundance$^{a}$} & N$_{\rm H}^{b}$ &Flux$^{c}$&
$\chi^{2}$/(dof)\\
         & (keV)& & N & O & Ne & S & Fe &  &  &  \\
\hline
\multicolumn{11}{c}{XMMU J004339.1+412654 ($\#3$)}\\
\hline
RS &$0.09^{+0.01}_{-0.01}$& ... &1 & 1 & 1 & 1 & 1 & $0.88^{+0.06}_{-0.08}$ & $0.26\pm0.02$&39.8(34)\\
   &$0.12^{+0.02}_{-0.01}$& ... &$0.98^{d}$&$0.44^{d}$&$1.00^{d}$&$0.57^{d}$&1&$0.61^{+0.08}_{-0.14}$&$0.26\pm0.02$&36.5(34)\\
   &$0.26^{+0.03}_{-0.04}$& ... &$0.98^{d}$&$0.19\pm0.07$&$0.64^{+0.30}_{-0.24}$&$0.57^{d}$&$0.15^{+0.08}_{-0.05}$&$0.11^{+0.09}_{-0.04}$&$0.29\pm0.02$&26.3(31)\\ 
\hline
NEI&$0.22^{+0.07}_{-0.04}$&$10.8^{+0.4}_{-0.2}$& 1 & 1 & 1 & 1 & 1 &$0.52^{+0.08}_{-0.03}$&$0.26\pm0.02$&46.0(33)\\
   &$1.06^{+0.98}_{-0.53}$&$9.9^{+0.2}_{-0.1}$&$0.98^{d}$&$0.44^{d}$&$1.00^{d}$&$0.57^{d}$&1&$0.14^{+0.09}_{-0.08}$&$0.29\pm0.02$&33.1(33)\\
\hline
\multicolumn{11}{c}{XMMU J004513.9+413614 ($\#21$)}\\
\hline
RS &$0.24\pm0.06$& ... &1 & 1 & 1 & 1 & 1 & $0.54^{+0.14}_{-0.18}$ & $0.13\pm0.01$&15.3(14)\\
   &$0.24^{+0.07}_{-0.06}$& ... &$0.61^{d}$&$0.53^{d}$&$1.00^{d}$&$0.42^{d}$&1&$0.47^{+0.13}_{-0.25}$&$0.12\pm0.01$&19.4(14)\\
   &$0.45^{+0.10}_{-0.04}$& ... &$0.61^{d}$&$0.53^{d}$&$1.00^{d}$&$0.42^{d}$&$0.18^{+0.09}_{-0.06}$&$0.09^{+0.08}_{-0.06}$&$0.16\pm0.01$&9.8(13)\\
\hline
\multicolumn{11}{c}{XMMU J004451.1+412907 ($\#17$)}\\
\hline
RS &$0.67^{+0.60}_{-0.18}$& ... &$<0.01$ & $<0.01$ & $<0.01$ & $<0.01$ & $<0.01$ & $0.07^{+0.04}_{-0.06}$ & $0.12\pm0.01$&7.9(10)\\
\hline
\end{tabular}

\begin{list}{}{}
\item $^{a}$ -- Relative to the solar abundance.
\item $^{b}$ -- An equivalent hydrogen column density in units of $10^{22}$ cm$^{-2}$.
\item $^{c}$ -- Absorbed model flux in the $0.3 - 3.0$ keV energy range in units of 
$10^{-13}$ erg s$^{-1}$ cm$^{-2}$.
\item $^{d}$ -- Fixed at optical value (\cite{Blair82}).
\end{list}
\end{table}

\begin{table}
\small
\caption{Bright Galactic foreground stars absorbed RS model spectral fit results ({\em XMM-Newton}/EPIC data, $0.3 - 3.0$ keV 
energy range). \label{spec_par_FRGS}}
\small
\begin{tabular}{ccccccc}
\hline
\hline
Source ID & $kT$   &A$_{\rm Fe}$& N$_{\rm H}$               &Flux$^{a}$&$\chi^{2}$& Remarks\\
          & (keV)  &            & $\times 10^{21}$ cm$^{-2}$&          & (dof)    &        \\
\hline
 6&$0.84^{+0.05}_{-0.07}$&$0.16^{+0.09}_{-0.06}$&$0.34^{b}$&$0.21\pm0.02$&$19.5(19)$&pn\\
15&$0.79^{+0.04}_{-0.06}$&$0.30^{+0.09}_{-0.07}$&$0.31^{b}$&$0.22\pm0.01$&$45.6(24)$&pn\\
20&$0.80^{+0.05}_{-0.08}$&$0.13^{+0.06}_{-0.05}$&$0.12^{b}$&$0.50\pm0.02$&$34.9(22)$&pn\\
22&$0.66^{+0.04}_{-0.03}$&$0.17^{+0.03}_{-0.07}$&$0.06^{b}$&$1.24\pm0.06$&$58.8(29)$&pn, Obs. $\#2$\\
22&$0.65\pm0.02$&$0.10\pm0.03$&$0.02^{b}$&$2.70\pm0.06$&$149.3(91)$&pn+MOS1, Obs. $\#3$\\
36&$0.66^{+0.03}_{-0.04}$&$0.57^{+0.15}_{-0.11}$&$0.27^{+0.40}_{-0.27}$&$0.32\pm0.02$&$31.9(33)$&pn+MOS1+MOS2\\
\hline
\end{tabular}

\begin{list}{}{}
\item $^{a}$ -- absorbed model flux in the $0.3 - 3.0$ keV energy range in 
units of $10^{-13}$ erg s$^{-1}$ cm$^{-2}$
\item $^{b}$ -- a $2\sigma$ upper limit
\end{list}
\end{table}

\begin{table}
\small
\caption{Radio sources spectral fit results ({\em XMM-Newton}/EPIC data, absorbed simple power law model, $0.3 - 10.0$ keV energy range). \label{spec_par_radio}}
\small
\begin{tabular}{cccccc}
\hline
\hline
Source ID & Photon   & N$_{\rm H}$               &Flux$^{a}$&$\chi^{2}$& Remarks\\
          & Index    & $\times 10^{22}$ cm$^{-2}$&          & (dof)    &        \\
\hline
16 &$2.15^{+0.11}_{-0.13}$&$0.17^{+0.03}_{-0.04}$&$0.73\pm0.06$&19.1(25)& pn\\
28 &$1.86^{+0.10}_{-0.12}$&$0.40^{+0.10}_{-0.05}$&$1.00\pm0.04$&61.3(50)& pn+MOS1+MOS2\\
\hline
\end{tabular}

\begin{list}{}{}
\item $^{a}$ -- absorbed model flux in the $0.3 - 10$ keV energy range in 
units of $10^{-13}$ erg s$^{-1}$ cm$^{-2}$
\end{list}
\end{table}

\begin{table}
\small
\caption{Other bright X-ray sources spectral fit results ({\em XMM-Newton}/EPIC data, absorbed simple power law model, 
$0.3 - 10$ keV energy range). \label{spec_par_other}}
\small
\begin{tabular}{cccccl}
\hline
\hline
Source Name   & Photon   & N$_{\rm H}$               &Flux$^{a}$&$\chi^{2}$& Remarks\\
XMMU J00...   & Index    & $\times 10^{22}$ cm$^{-2}$&          & (dof)    &        \\
\hline
 8 &$1.62^{+0.13}_{-0.09}$&$0.06^{+0.02}_{-0.04}$&$0.62\pm0.03$&47.5(44)  & pn \\
 9 &$2.88^{+0.40}_{-0.30}$&$0.72^{+0.21}_{-0.23}$&$0.19\pm0.02$&13.7(17)  & pn \\
10 &$1.91^{+0.13}_{-0.15}$&$0.12^{+0.04}_{-0.05}$&$0.33\pm0.02$& 9.6(20)  & pn \\
13 &$1.85\pm0.10$         &$0.06\pm0.03$         &$1.65\pm0.07$&23.3(35)  & MOS1+MOS2 \\
14 &$1.97\pm0.07$         &$0.19^{+0.02}_{-0.03}$&$1.16\pm0.04$&60.4(57)  & pn+MOS1+MOS2\\
18 &$1.79^{+0.17}_{-0.17}$&$0.10^{+0.04}_{-0.06}$&$1.40\pm0.10$& 5.9(11)  & MOS2 \\
19 &$1.55^{+0.21}_{-0.16}$&$0.47^{+0.19}_{-0.11}$&$1.72\pm0.11$&31.2(43)  & MOS1+MOS2 \\
23 &$2.05^{+0.23}_{-0.18}$&$0.30^{+0.08}_{-0.06}$&$0.54\pm0.04$&12.6(13)  & pn \\
24 &$1.67^{+0.08}_{-0.07}$&$0.25^{+0.03}_{-0.04}$&$1.19\pm0.05$&40.1(45)  & pn+MOS1+MOS2\\
25 &$1.61^{+0.14}_{-0.13}$&$0.18^{+0.08}_{-0.06}$&$0.74\pm0.04$&22.9(28)  & MOS1+MOS2 \\
29 &$1.78^{+0.30}_{-0.24}$&$0.30^{+0.16}_{-0.09}$&$0.34\pm0.03$&10.6(14)  & pn \\
30 &$1.54\pm0.29$         &$0.10^{+0.08}_{-0.05}$&$0.27\pm0.02$&13.3(8)   & pn \\
31 &$2.29^{+0.05}_{-0.04}$&$0.15\pm0.01$         &$4.75\pm0.06$&319.5(295)& pn+MOS1+MOS2\\
32 &$1.79^{+0.23}_{-0.19}$&$0.23\pm0.06$         &$0.91\pm0.08$&10.1(13)  & pn\\
33 &$1.39^{+0.38}_{-0.22}$&$0.16^{+0.21}_{-0.08}$&$0.67\pm0.06$&23.8(14)& pn \\
34 &$1.94^{+0.09}_{-0.08}$&$0.10\pm0.03$         &$0.87\pm0.03$&55.1(58  )& pn+MOS1+MOS2 \\
35 &$1.61^{+0.17}_{-0.18}$&$0.10\pm0.06$         &$0.56\pm0.05$&12.8(15)  & pn \\
37 &$1.73^{+0.07}_{-0.08}$&$0.23^{+0.02}_{-0.03}$&$5.83\pm0.14$&76.4(71)  & MOS1+MOS2\\
\hline
\end{tabular}

\begin{list}{}{}
\item $^{a}$ -- absorbed model flux in the $0.3 - 10$ keV energy range in 
units of $10^{-13}$ erg s$^{-1}$ cm$^{-2}$
\end{list}
\end{table}

\clearpage

\clearpage

\clearpage

\clearpage

\clearpage

\clearpage

\clearpage

\clearpage

\clearpage

\clearpage

\clearpage

\clearpage

\clearpage

\begin{thebibliography}{}
\bibitem[Anders \& Grevesse 1989]{AG89}Anders, E., \& Grevesse, N. 1989, Geochim. 
Cosmochim. Acta, 53, 197
\bibitem[Arnaud 1996]{arnaud96}Arnaud, K. 1996, in Astronomical Data Analysis 
Software and Systems V, ASP Conference Series 101, ed. G. Jacoby \& J. Barnes 
(San Francisco: ASP) 17
\bibitem[Barmby et al. 2000]{Barmby00}Barmby, P., Huchra, J.P., Brodie, J.P., 
Forbes, D.A., Schroder, L.L., \& Grillmar, C.J. 2000, AJ, 119, 727 
\bibitem[Barmby \& Huchra 2001]{Barmby01}Barmby, P., \& Huchra, J.P. 2001,
ApJ, 122, 2458
\bibitem[Battistini et al. 1987]{Bo87}Battistini, P., et al., 1987, A$\&$AS, 
67, 447
\bibitem[Blair et al. 1981]{Blair81}Blair, W. P., Kirshner, R. P., $\&$ 
Chevalier, R. A. 1981, ApJ, 247, 879
\bibitem[Blair et al. 1982]{Blair82}Blair, W. P., Kirshner, R. P., $\&$ 
Chevalier, R. A. 1982, ApJ, 254, 50
\bibitem[Condon et al. 1998]{nvss}Condon, J.J., Cotton, W.D., Greisen, E.W., 
Yin, Q.F., Perley, R.A., Taylor, G.B., \& Broderick J.J. 1998, AJ, 115, 1693
\bibitem[DiSalvo et al. 2001]{DiSalvo01}DiSalvo, T., Robba, N. R., Iaria, R., 
Stella, L., Burderi, L., \& Israel, G. L. 2001, ApJ, 554, 49
\bibitem[Di Stefano et al. 2001]{DiStefano01}Di Stefano, R., Kong, A. K. H., 
Garcia, M. R., Barmby, P., Greiner, J., Murray, S. S., $\&$ Primini, F. A. 
2002, ApJ, 570, 618
\bibitem[Di Stefano \& Kong 2002]{DiStefano02}Di Stefano, R., \& Kong, A. K. H. 
2002, ApJ, 592, 884
\bibitem[Dickey \& Lockman 1990]{DL90}Dickey, J. M., \& Lockman F. J. 1990, 
ARA$\&$A, 28, 215
\bibitem[Dove et al. 1997]{Dove97}Dove, J. B., Wilms, J., Maisack, M., $\&$ 
Begelman, M. C. 1997, ApJ, 487, 759
\bibitem[Garcia et al. 2002]{chandra_circ02}Garcia, M. R., Kong, A. K. H., 
McClintock, J. E., Primini, F. A., Kaaret, P., \& Murray, S. S. 2002, ATel. 82 
\bibitem[Haas et al. 1998]{Haas98}Haas, M., Lemke, D., Stickel, M., Hippelein, 
H., Kunkel, M., Herbstmeier, U., \& Mattila, K. 1998, A$\&$A, 338, L33
\bibitem[Haberl et al. 2003]{HDP03}Haberl, F., Dennerl, K., \& Pietsch, W. 
2003, A$\&$A, 406, 471
\bibitem[Haiman et al. 1994]{Haiman94}Haiman, Z., Magnier, E. A., Lester, R. R., 
Lewin, W. H. G., van Paradijs, J., Hasinger, G., Pietsch, W., \& Truemper, J. 
1994, A$\&$A, 286, 725
\bibitem[Huchra, Brodie \& Kent 1991]{Huchra91}Huchra, J.P., Brodie, J.P., \& 
Kent, S.M. 1991, ApJ, 370, 495
\bibitem[Iaria et al. 2001]{Iaria01}Iaria, R., Burderi, L., Di Salvo, T., 
La Barbera, A., \& Robba, N. R. 2001, ApJ, 547, 412
\bibitem[Kahabka \& van den Heuvel 1997]{KVDH97} Kahabka, P., \& van den 
Heuvel, E.~P.~J. 1997, ARA$\&$A, 35, 69
\bibitem[Kong et al. 2002]{Kong02}Kong, A. K. H., Garcia, M. R., Primini, F. A., 
Murray, S. S., Di Stefano, R., \& McClintock, J. E. 2002, ApJ, 577, 738
\bibitem[Kotov, Trudolyubov \& Vestrand 2003]{kotov03}Kotov, O., Trudolyubov, S., 
\& Vestrand, W. T. 2003, ApJ, submitted
\bibitem[Long et al. 1996]{Long96}Long, K., Charles, P., Blair, W., \& Gordon, S. 
1996, ApJ, 466, 750
\bibitem[Magnier et al. 1992]{Magnier92}Magnier, E. A., Lewin, W. H. G., van 
Paradijs, J., Hasinger, G., Jain, A., Pietsch, W., \& Truemper, J. 1992, 
A$\&$AS, 96, 379
\bibitem[Magnier 1993]{Magnier93}Magnier, E. A. 1993, Ph.D. thesis, MIT
\bibitem[Magnier et al. 1995]{Magnier95}Magnier, E. A., Prins, S., van Paradijs, J., 
Lewin, W. H. G., Supper, R., Hasinger, G., Pietsch, W., Truemper, J. 1995, A$\&$AS, 
114, 215
\bibitem[Magnier et al. 1997]{Magnier97}Magnier, E. A., Primini, F. A., Prins, S., 
van Paradijs, J., \& Lewin, W. H. G. 1997, ApJ, 490, 649
\bibitem[Massey et al. 2001]{Massey01}Massey, P., Hodge, P. W., Holmes, S., 
Jacoby, G., King, N. L., Olsen, K., Saha, A., \& Smith, C. 2001, in 
American Astronomical Society Meeting, 199, 1305 
\bibitem[Mitsuda et al. 1984]{Mitsuda84}Mitsuda, K., Inoue, H., Koyama, K., et al. 
1984, PASJ, 36, 741
\bibitem[Perrett et al. 2002]{Perrett02}Perrett, K.M., Bridges, T.J., Hanes, D.A., 
Irwin, M.J., Brodie, J.P., Carter, D., Huchra, J.P., \& Watson, F.G. 2002, AJ, 
123, 2490
\bibitem[Pietsch et al. 2003]{Pietsch03}Pietsch, W., Ehle, M., Haberl, F., 
Misanovic, Z., Trinchieri, G. 2003, Astronomische Nachrichten, 324, 85 
\bibitem[Prestwich et al. 2003]{Prestwich03}Prestwich, A. H., Irwin, J. A., 
Kilgard, R. E., Krauss, M. I., Zezas, A., Primini, F., Kaaret, P., \& 
Boroson, B. 2003, ApJ, 595, 719
\bibitem[Primini et al. 1993]{Primini93}Primini, F. A., Forman, W., \& 
Jones, C., 1993, ApJ, 410, 615
\bibitem[Ramsay et al. 1994]{Ramsay94}Ramsay, G., Mason, K. O., Cropper, M., 
Watson, M. G., \& Clayton, K. L. 1994, MNRAS, 270, 692
\bibitem[Raymond \& Smith 1977]{RS}Raymond, J. S., \& Smith, B. W. 1977, 
ApJS, 35, 419
\bibitem[Schmidtobreick, Haas \& Lemke 2000]{schmidtobreick00}Schmidtobreick, 
L., Haas, M., \& Lemke, D. 2000, A$\&$A, 363, 917
\bibitem[Shirey et al. 2001]{Shirey01}Shirey, R., Soria, R., Borozdin, K., 
Osborne, J. P., Tiengo, A., Guainazzi, M., Hayter, C., La Palombara, N., 
Mason, K., Molendi, S., Paerels, F., Pietsch, W., Priedhorsky, W., Read, 
A. M., Watson, M. G., West, R. G., 2001, A$\&$A, 365, L195
\bibitem[Sidoli et al. 2001]{Sidoli01}Sidoli, L., Parmar, A. N., Oosterbroek, T., 
Stella, L., Verbunt, F., Masetti, N., \& Dal Fiume, D., 2001, A$\&$A, 368, 451
\bibitem[Strueder et al. 2001]{Strueder01}Strueder, L. et al., 2001, A\&A, 
L18
\bibitem[Sunyaev $\&$ Titarchuk 1980]{ST80}Sunyaev, R. A., $\&$ Titarchuk, 
L. G. 1980, A$\&$A, 86, 121
\bibitem[Supper et al. 1997]{Supper97}Supper, R., Hasinger, G., Pietsch, W., 
Truemper, J., Jain, A., Magnier, E.A., Lewin, W.H.G., $\&$ van Paradijs, J. 
1997, A$\&$A, 317, 328  
\bibitem[Supper et al. 2001]{Supper01}Supper, R., Hasinger, G., Lewin, W. H. G., 
Magnier, E.A., van Paradijs, J., Pietsch, W., Read, A.M., \& Truemper, J. 2001, 
A$\&$A, 373, 63 
\bibitem[Titarchuk 1994]{T94}Titarchuk, L. 1994, ApJ, 434, 570
\bibitem[Titarchuk \& Lyubarskij 1995]{TL95}Titarchuk, L., \& Lyubarskij, Y. 
1995, ApJ, 450, 876
\bibitem[Trinchieri \& Fabbiano 1991]{TF91}Trinchieri, G., \& Fabbiano, G., 
1991, ApJ, 382, 82
\bibitem[Trudolyubov et al. 2002a]{xmm_circ02}Trudolyubov, S., Borozdin, K., \& 
Priedhorsky, W., Mason, K., \& Cordova, F. 2002, IAU Circ. 7798
\bibitem[Trudolyubov et al. 2002b]{dipper02}Trudolyubov, S., Borozdin, K., \& 
Priedhorsky, W., Osborne, J., Watson, M., Mason, K., \& Cordova, F. 2002b, ApJ, 
581, L27
\bibitem[Turner et al. 2001]{Turner01}Turner, M. et al., 2001, A\&A, 365, 
L27
\bibitem[Vaughan et al. 1994]{Vaughan94}Vaughan, B. A., van der Klis, M., Wood, K. S., 
Norris,  J. P., Hertz, P., Michelson, P. F., van Paradijs, J., Lewin, W. H. G., Mitsuda, 
K., \& Penninx, W. 1994, ApJ, 435, 362 
\bibitem[West, Barber, \& Folgheraiter 1997]{West97}West, R. G., Barber, C. R., \& 
Folgheraiter, E. L. 1997, MNRAS, 287, 10
\bibitem[Williams et al. 2003]{Williams03}Williams, B. F., Garcia, M. R., Kong, A. K. H., 
Primini, F. A., King, A. R., \& Murray, S. S. 2003, ApJ, submitted, astro-ph/0306421
\end{thebibliography}
\end{document}